\documentclass[
	aps, prd,  notitlepage, 
        floats, floatfix, onecolumn,
	amsmath, amssymb, amsfonts, eqsecnum,
	superscriptaddress,
	showpacs, showkeys,
	nofootinbib,
 	longbibliography,
]{revtex4-1}
\usepackage{multirow}
\usepackage{epsfig}
\usepackage{graphicx}
\usepackage{float}
\usepackage{subfig}
\usepackage[usenames,dvipsnames]{xcolor}
\usepackage{tikz}
\usepackage{xspace} 
\usepackage{bm}
\usepackage[utf8]{inputenc} 
\usepackage{latexsym}
\usepackage{ulem}
\renewcommand{\emph}[1]{\textit{#1}}
\xdefinecolor{mylinkcolor}{rgb}{0,0,0.5}
\usepackage[
	bookmarksnumbered, bookmarksopen, bookmarksopenlevel=2,
	breaklinks=true,
	colorlinks=true, filecolor=mylinkcolor, citecolor=mylinkcolor,
	linkcolor=mylinkcolor, urlcolor=mylinkcolor, menucolor=mylinkcolor,
]{hyperref}
\DeclareMathAlphabet{\mathsfsl}{OT1}{cmss}{m}{sl}

\newcommand{\rmd}{\mathrm{d}}
\def\be{\begin{equation}}
\def\ee{\end{equation}}
\def\bea{\begin{eqnarray}}
\def\eea{\end{eqnarray}}
\newcommand{\bes}{\begin{subequations}}
\newcommand{\ees}{\end{subequations}}
\def\comment#1{}

\newcommand{\RNum}[1]{\uppercase\expandafter{\romannumeral #1\relax}}

\def\msun{\,{\rm M_\odot}}

\usepackage{color}

\allowdisplaybreaks

\begin{document}

\title{Geometrized effective-one-body formalism for extreme-mass-ratio limits: Generic orbits}
\date{\today}

\author{Chen Zhang}
\affiliation{Shanghai Astronomical Observatory, Chinese Academy of Sciences, Shanghai 200030, P. R. China}
\affiliation{School of Astronomy and Space Science, University of Chinese Academy of Sciences Beijing, 100049, P. R. China}
\author{Wen-Biao Han}
\email{Corresponding author: wbhan@shao.ac.cn}
\affiliation{Shanghai Astronomical Observatory, Chinese Academy of Sciences, Shanghai 200030, P. R. China}
\affiliation{School of Fundamental Physics and Mathematical Sciences, Hangzhou Institute for Advanced Study, UCAS, Hangzhou 310024, China}
\affiliation{School of Astronomy and Space Science, University of Chinese Academy of Sciences Beijing, 100049, P. R. China}
\affiliation{International Centre for Theoretical Physics Asia-Pacific, Beijing/Hangzhou 310024, China}
\author{Xing-Yu Zhong}
\affiliation{Shanghai Astronomical Observatory, Chinese Academy of Sciences, Shanghai 200030, P. R. China}
\affiliation{School of Astronomy and Space Science, University of Chinese Academy of Sciences Beijing, 100049, P. R. China}
\author{Gang Wang}
\affiliation{Shanghai Astronomical Observatory, Chinese Academy of Sciences, Shanghai 200030, P. R. China}
\affiliation{School of Astronomy and Space Science, University of Chinese Academy of Sciences Beijing, 100049, P. R. China}

\begin{abstract}
Compact objects inspiraling into supermassive black holes, known as extreme-mass-ratio inspirals, are
an important source for future space-borne gravitational-wave detectors. When constructing waveform templates, usually the adiabatic approximation is employed to treat the compact object as a test particle for a short duration, and the radiation reaction is reflected in the changes of the constants of motion. However, the mass of the compact object should have contributions to the background. In the present paper, employing the effective-one-body formalism, we analytically calculate the trajectories of a compact object
around a massive Kerr black hole with generally three-dimensional orbits and express the fundamental
orbital frequencies in explicit forms. In addition, by constructing an approximate “constant” similar
to the Carter constant, we transfer the dynamical quantities such as energy, angular momentum, and the
“Carter constant” to the semilatus rectum, eccentricity, and orbital inclination with mass-ratio corrections. The linear mass-ratio terms in the formalism may not be sufficient for accurate waveforms, but our analytical method for solving the equations of motion could be useful in various approaches to building
waveform models.
\end{abstract}

\maketitle

\section{Introduction}
One of the most promising and rewarding sources of gravitational waves for low-frequency, space-based gravitational-wave (GW) detectors—such as the future Laser Interferometer Space Antenna(LISA)~\cite{lisa}, Taiji~\cite{hu2017taiji} and Tian-Qin~\cite{luo2016tianqin}—are the so-called extreme-mass-ratio inspirals (EMRIs)~\cite{amaro2007intermediate,babak2017science,berry2019unique}, which are compact objects (COs) such
as neutron stars or stellar-mass (or stellar origin) black holes (SOBHs) inspiraling into supermassive black holes (SMBHs) in the mass range $10^5$-$10^7\msun$.

The majority of observed EMRI events are expected to
be SOBH-SMBH mergers; this is partly due to mass
segregation concentrating heavier BHs in the Galactic
center and partly because their louder intrinsic amplitude enables them to be detected out to greater distances. The small CO is usually approximated as a test particle, and over the short orbital time scale follows a nearly geodesic trajectory in the background metric of the SMBH. The system radiates GWs at harmonics of the geodesic frequencies, so the GW frequency spectrum encodes details of the instantaneous geodesic trajectory.

Though the signals from EMRIs are usually very weak,
after one year of observation the signal-to-noise ratio may be enough to be detected ~\cite{amaro2007intermediate}. To detect this kind of long duration signals, the requirement on the accuracy of waveform templates is that the dephasing should be less than a few radians after $10^5$ cycles~\cite{gair2013testing,babak2017science}.

Nowadays, there are several kinds of EMRI templates. The first kind uses Teukolsky equations~\cite{Teukolsky} and treats COs as test particles (omitting the mass in their conservation dynamics part) and thus they just extract a snapshot waveform; however, accurate Teukolsky-based waveforms are computationally expensive to generate. Another kind uses the post-Newtonian (PN) fluxes approximation to describe the evolution of orbital parameters with gravitational radiation reaction, in particular the so-called kludge
waveforms~\cite{barack2004lisa,babak2007kludge,chua2017augmented}. The third kind considers the correction
due to the small mass by using the effective-one-body
(EOB) formalism, but does not consider the spin of the
small object, such as in circular orbits with PN waveforms~\cite{yunes2010modeling,yunes2011extreme}  
and the Teukolsky-based waveforms~\cite{han2011constructing,han2014gravitational}. The kludge models can generate the waveforms quickly for three-dimensional (3D) orbits, but they take the COs as test particles in the orbital motion even though they include the radiation reaction.

 If the time scale of GW radiation is much larger than that of the orbital motion, then at a given instant the properties of motion can be calculated from a conservative equation $\ddot{x}^\mu + ^B\Gamma^\mu_{\rho\sigma}\dot{x}^\mu\dot{x}^\nu = 0 + O(\nu) + O(\nu^2) + \cdots$. Here the label ``B" represents the background field of the massive body. In popular EMRI template models, like AAK~\cite{barack2004lisa}, NK~\cite{babak2007kludge}, AAK~\cite{chua2017augmented}, XSPEG~\cite{xin2019gravitational} etc, the mass-ratio corrections in the conservative dynamics are omitted, i.e., the right-hand
side of the equation is exactly zero. This is just the geodesic equation of a test particle around a massive black hole. The test particle–adiabatic approximation (adiabatic model) makes the problem much simpler, but induces errors in the GW simulation. As in Eq.(74) of~Ref.\cite{barack2018self}, the mass ratio expansion to the phase evolution is
\be
\phi(t) = \frac{1}{\nu}  [\phi_{(0)}+ \nu \phi_{(1)}+ O(\nu^2)].
\ee
An EMRI model that captures the leading-order term $\phi_{(0)}$ is called an
adiabatic model. The calculation of $\phi_{(0)}$  requires only the averaged dissipative first-order self-force (in the mass-ratio expansion) that is fully equivalent to the fluxes, and some literature states that in general it may be accurate enough for detecting EMRI signals using LISA~\cite{hughes2018adiabatic}. The next-to-leading-order term, which includes both the dissipative and conservative
pieces of the first-order self-force as well as the averaged dissipative piece of the second-order self-force. The post-1 adiabatic phase is required for the parameter estimation of EMRIs~\cite{hughes2018adiabatic}. In the present paper, we mainly focus on the conservation dynamics part of EMRI, and try to analytically solve the equation of motion which includes the first-order mass-ratio corrections. The mass-ratio terms we use may not
be enough for final EMRI waveforms, so the aim of this
paper is not to construct the waveform template, but rather to demonstrate the method for the analytical solution of orbits. Our method should be useful once there are some more precise self-force corrections.

The EOB formalism, by including the mass-ratio corrections up to a certain order in the PN expansion, can well describe the dynamical evolution of binary black holes~\cite{buonanno1999eff,buonanno2000transition}, and is widely used to construct the waveform templates for LIGO~\cite{taracchini2014effective,buonanno2007approaching,purrer2016frequency,husa2016frequency,khan2016frequency,chu2016accuracy,kumar2016accuracy,pan2014inspiral}. Most of these models only consider the circular-orbit cases. An eccentric EOB numerical relativity waveform template for spinning black holes was developed ~\cite{cao2017waveform}, but the orbits were not geometrized and the orbital parameters were not well defined. Recently, an analytically eccentric EOB formalism for Schwarzschild BHs was given~\cite{hinderer2017foundations}. In a previous work~\cite{zc2020}, we present an analytically equatorial-eccentric EOB formalism for spinning cases in the extreme-mass-ratio limit. Note that
the EOB formalism’s correction in the extreme-mass-ratio
limit has not been guaranteed. However, as stated in~Ref.\cite{albanesi2021effective}, the extreme-mass-ratio limit plays a pivotal role in the EOB development, especially for what concerns waveforms and fluxes, which can be informed by and compared with numerical results ~\cite{Nagar:2006xv,Damour:2007xr,Bernuzzi:2010ty,Bernuzzi:2011bc,Bernuzzi:2010xj,Bernuzzi:2011aj,Harms:2014an,Nagar:2014kha,Harms:2015ixa,Harms:2016ctx,Lukes:2017st,Nagar:2019wrt}. In addition, the EOB formalism with the extreme-mass-ratio limit should be improved due to a lot of works dedicated to analytically calculating gravitational self-force terms and providing comparisons with numerical results ~\cite{Damour:2009sm,Barack:2010ny,Akcay:2012ea,Bini:2014ica,Bini:2015mza,Akcay:2015pjz,Barack:2019agd}.

It is well known that the orbits of EMRIs could be highly eccentric~\cite{babak2017science} with orbital inclination (orbital plane procession), and the supermassive black hole in the center should be spinning in general. In the present paper, we extend the previous work by Hinderer and Babak~\cite{hinderer2017foundations} and ourselves ~\cite{zc2020} to the inclined-eccentric orbits in the Kerr background with the extreme-mass-ratio limit. We analytically transfer the original EOB dynamical equations to geometric kinetic motion with the semilatus rectum $p$, eccentricity $e$ and orbital inclination $\iota$ as the orbital parameters, together with
three phase variables associated with the spatial geometry of the radial, azimuthal, and polar motion denoted by $(\xi,\phi,\chi)$. Because of the extremely small mass ratio,we omit the spin of
the effective small body, and thus the very complicated spin-spin coupling terms disappear but maintain enough accuracy. 

An important feature of the dynamics of an extrememass-ratio binary system in a bounded inclined-eccentric
orbit is that the orbit can be characterized by three
frequencies: the radial frequency $\omega_r$ associated with the libration between the apoapsis and periapsis, the polar frequency  $\omega_\theta$ associated with the libration between $\theta_{min}$ and $\pi-\theta_{min}$, and the azimuthal rotational frequency $\omega_\phi$. Once these three frequencies and orbital parameters are obtained, the frequency of GWs can be obtained and may be encoded in the Teukolsky equation to get accurate waveforms~\cite{han2010gravitational,han2011constructing,han2014gravitational,han2017excitation,cai2016gravitational,yang2019testing,cheng2019highly}. 

The organization of this paper is as follows. The basic knowledge of EOB formalism is introduced in the following section. In Sec.~\ref{sec:geometrization}, we reparametrize the original
spinning EOB dynamical description to a geometric formalism in the more efficient reparametrized terms of $(p,~e,~\iota,~\xi,~\chi,~\phi)$. We analytically express the fundamental
frequencies in three integrals with two parameters: $\xi$ and $\chi$. In particular, we investigate the influence of the mass ratio on the detection of EMRIs. We also give the general forms for the evolution of orbital parameters with gravitational radiation reaction. Finally, the last section contains our conclusions and discussions.

Throughout this paper we use geometric units $G = c = 1$, the units of time and length are the mass of
system $M$, and the units of linear and angular momentum are $\mu$ and $\mu M$, respectively, where $\mu$ is the reduced mass of the effective body.

\section{Effective-one-body Hamiltonian}
\label{sec:HEOB}
 The EOB formalism was originally introduced in Refs.~\cite{buonanno1999eff,buonanno2000transition} to describe the evolution of a binary system. We start by considering an EMRI system with a central Kerr black hole $m_1$ and inspiraling object $m_2$ (we assume
that it is nonspinning for simplicity, $m_2 \ll m_1$). For the moment, we neglect the radiation-reaction effects and focus on purely geodesic motion. The conservative orbital dynamics is derived via Hamilton's equations using the EOB Hamiltonian 
 \be
 H_{\rm EOB}=M\sqrt{1+2\nu (\hat{H}_{\rm eff}-1)}\,,\label{eq:HEOB}
 \ee
where $M=m_1+m_2$, $\nu=m_1m_2/M^2$, $\mu=\nu M$ and the reduced effective Hamiltonian $\hat{H}_{\rm eff}=H_{\rm eff}/\mu$. The deformed Kerr metric is given by~\cite{barausse2010improved}
\bes
\bea
\label{def_metric_in}
g^{tt} &=& -\frac{\Lambda_t}{\Delta_t\,\Sigma}\,,\\
g^{rr} &=& \frac{\Delta_r}{\Sigma}\,,\\
g^{\theta\theta} &=& \frac{1}{\Sigma}\,,\\
g^{\phi\phi} &=& \frac{1}{\Lambda_t}
\left(-\frac{\widetilde{\omega}_{\rm fd}^2}{\Delta_t\,\Sigma}+\frac{\Sigma}{\sin^2\theta}\right)\,,\label{eq:gff}\\
g^{t\phi}&=&-\frac{\widetilde{\omega}_{\rm fd}}{\Delta_t\,\Sigma}\,.\label{def_metric_fin}
\eea
\ees
The quantities $\Sigma$, $\Delta_t$, $\Delta_r$, $\Lambda_t$, and $\widetilde{\omega}_{\rm fd}$
in Eqs.~(\ref{def_metric_in})-(\ref{def_metric_fin}) are given by 
\bes
\bea
\Sigma &=& r^2+a^2 \cos ^2\theta \,, \\
\label{deltat}
\Delta_t &=& r^2\, \left [A(u) + \frac{a^2}{M^2}\,u^2 \right ]\,, \\
\label{deltar}
\Delta_r &=& \Delta_t\, D^{-1}(u)\,,\\
\Lambda_t &=& \left(r^2+a^2\right)^2-a^2\Delta _{t}\sin ^2\theta \,,\\
\widetilde{\omega}_{\rm fd} &=& 2 a\, M\, r+ \omega_1^{\rm fd}\,\nu\,\frac{a M^3}{r}+ \omega_2^{\rm fd}\,\nu\,\frac{M a^3}{r}
\label{eq:omegaTilde}\,,
\eea
\ees
where $a=|\bm{S}_{\rm{Kerr}}|/M$ is the effective Kerr parameter and $u = M/r$. The values of $\omega_1^{\rm fd}$ and $\omega_2^{\rm fd}$ given by a preliminary comparison of EOB model with numerical relativity results are about $-10$ and $20$~\cite{rezzolla2008final,barausse2009predicting}, and the metric potentials $A$ and $D$ for the EOB model are given as follows.

The log-resummed, calibrated A potential is given by the expression from APPENDIX A of Ref.~\cite{steinhoff2016Apotential}
\begin{widetext}
\bea
A(u)&=&\Delta_u-\frac{a^2}{M^2}u^2,\\
\Delta_u&=&\bar{\Delta}_u \left(\Delta _0 \nu +\nu 
   \log \left(\Delta_5 u^5+\Delta_4
   u^4+\Delta_3 u^3+\Delta_2
   u^2+\Delta_1 u+1\right)+1\right),
   \eea
with
\bes
\begin{align}
\bar{\Delta}_u&=\frac{a^2u^2}{M^2}+\frac{1}{(K \nu -1)^2}+\frac{2u}{K \nu -1},\\
\Delta_5&= (K \nu -1)^2 \bigg[ \frac{64}{5} \log  (u)
  + \left(-\frac{1}{3} a^2
  \left(\Delta_1^3-3\Delta_1
  \Delta_2+3 \Delta _3\right)\right. \nonumber \\
  &-\frac{\Delta_1^5-5\Delta_1^3
  \Delta_2+5\Delta_1^2 \Delta_3+5\Delta_1\Delta_2^2-5
  \Delta_2 \Delta_3-5\Delta_4\Delta_1}{5 (K \nu -1)^2}\nonumber \\
  &\left.+\frac{\Delta_1^4-4 \Delta _1^2\Delta_2+4\Delta_1\Delta_3+2\Delta_2^2-4 \Delta_4}{2 K \nu  -2}+\frac{2275 \pi^2}{512}+\frac{128 \gamma }{5}-\frac{4237}{60}+\frac{256
  \log (2)}{5}\right) \bigg] , \\
\Delta_4&=\frac{1}{96} \bigg[8 \left(6 a^2 \left(\Delta_1^2-2 \Delta_2\right) (K \nu -1)^2+3 \Delta_1^4+\Delta_1^3 (8-8 K \nu ) \right. \nonumber \\
&\left.-12 \Delta_1^2\Delta_2+12\Delta_1 (2
  \Delta_2 K \nu -2 \Delta_2+\Delta_3)\right)\nonumber\\
  &+48\Delta_2^2-64 (K \nu -1) (3 \Delta_3-47 K \nu +47)-123 \pi ^2 (K\nu -1)^2\bigg],\\
\Delta_3&=-a^2\Delta_1 (K \nu -1)^2-\frac{\Delta_1^3}{3}+\Delta_1^2 (K \nu -1)+\Delta_1\Delta_2-2 (K \nu -1) (\Delta_2-K
   \nu +1),\\
\Delta_2 &=\frac{1}{2} \left(\Delta_1 (\Delta_1-4 K
   \nu +4)-2 a^2\Delta_0 (K \nu -1)^2\right),\\
\Delta_1 &=-2 (\Delta_0+K) (K \nu -1),\\
\Delta_0&=K (K \nu -2),
\end{align}
where $K$ is a calibration parameter tuned to numerical-relativity simulations whose most recently updated value was determined in Eq.~(4.8) of Ref.~\cite{bohe2017improved}
\begin{align}
K=267.788 \nu ^3-126.687 \nu ^2+10.2573 \nu +1.7336
\end{align}
\ees
\end{widetext}

The $D$ potential is 
\bes
\begin{align}
D^{-1}(u)&=1+\log\left[D_{\rm Taylor}\right]\nonumber\\
D_{\rm Taylor}&=1+6\nu u^2+2 \nu u^3(26-3\nu).
\end{align}
\ees

\section{Geometrization of the conservative EOB dynamics}
\label{sec:geometrization}
The EOB Hamiltonian in Eq.~\eqref{eq:HEOB} is canonically transformed and subsequently mapped to an effective Hamiltonian $H_{\rm eff}$ describing a particle of effective mass $\mu$ and effective spin $S_*=a M(m_2/m_1)$, moving in a deformed Kerr metric of mass $M$ and effective spin $S_{\rm Kerr}=a M$, and the effective Hamiltonian is given by~\cite{barausse2010improved,Barausse2011Extending,Taracchini2012Prototype}
\be
H_{\rm eff}=H_{\rm NS}+H_{\rm S}-\frac{\nu}{2r^3} S_*^2 \,,\label{eq:Heff}
\ee
where the first part is just the Hamiltonian of a nonspinning particle in the deformed-Kerr metric with the mass-ratio correction. When $\nu \rightarrow 0$, this part goes back to the test-particle limit in Kerr spacetime. The forms of $H_s$ and $S_*$ are quite trivial; readers can refer to  \cite{Barausse2011Extending} for details. For EMRIs, due to the small mass-ratio $\nu\sim 10^{-7}\sim 10^{-4}$, as we have shown in~\cite{zc2020},  the last two terms in Eq.(\ref{eq:Heff}) can be neglected without loss of accuracy because they are two orders less than the mass ratio. The Hamiltonian equations for the orbital motion are
\begin{align}
\frac{d \boldsymbol{r}}{d {t}}  = \frac{\partial {H}_{\mathrm{EOB}}}{\partial \boldsymbol{P}}\,, \qquad
\frac{d \boldsymbol{P}}{d {t}}  = -\frac{\partial {H}_{\mathrm{EOB}}}{\partial \boldsymbol{r}} \label{eq:hamilton}\,.
\end{align}
In form, the above differential equations are coupling each other, especially for radial  $r$ and polar $\theta$ equations. Though the numerical integral can give accurate trajectories, analytical
ones with geometrized parameters will be valuable for
revealing the properties of motion. In the section, we follow the procedure of the test-particle case, decouple the motion in the  $r$ and $\theta$ directions, and give a geometrized formalism to replace the dynamical equation(\ref{eq:hamilton}).

\subsection{Semi-Carter ``constant" in deformed Kerr spacetime}
\label{sec:Carter constant}

The effective one-body dynamics was given by a Hamilton-Jacobi equation of the form~\cite{damour2001coalescence}
\be
\label{eq:superhamiltonian}
g^{\alpha \beta}P_\alpha P_\beta+\frac{Q_4 M^2 P_r^4}{r^2 \mu^2}+\mu^2=0\,.
\ee
 The function ${Q}_4=2(4-3\nu)\nu $~\cite{damour2000determination} represents a nongeodesic term that appears at 3PN order. We omit this term in the following calculations because it is  next-to-leading order in the mass ratio and the above equation cannot be separated if we retain this term. We use the deformed Kerr metric components~\eqref{def_metric_in}-\eqref{def_metric_fin} to bring Eq.~\eqref{eq:superhamiltonian} into the concrete form
\begin{widetext}
\bea
-\mu ^2\Sigma&=&-\frac{\left(r^2+a^2\right)^2-a^2 \sin ^2\theta (r^2A(u)+a^2)}{r^2A(u)+a^2}P_t^2 -2\frac{\widetilde{\omega}_{\rm fd}}{r^2A(u)+a^2}P_t P_{\phi } +  \Delta_r P_r^2  \nonumber\\
&&+P_{\theta }^2+\frac{1}{\left(r^2+a^2\right)^2-a^2 \sin ^2\theta  (r^2A(u)+a^2)}\left(\frac{(r^2+a^2 \cos ^2\theta)^2}{\sin ^2\theta }-\frac{{\widetilde{\omega}_{\rm fd}}^2}{r^2A(u)+a^2 }\right)P_{\phi }^2\label{eq: separation1}\,,
\eea
\end{widetext}
where $P_r$, $ P_{\phi }$, and $P_{\theta }$ are the canonical radial, azimuthal, and polar
angular momentum. From the symmetries we immediately obtain two constants of motion corresponding to the  conservation of energy, $H_{\rm eff}$, and angular momentum about the symmetry axis, $L_{z}$; thus, we have
\bes
\bea
P_t &=&-H_{\rm eff} \,, \label{eq:pt}\\
P_{\phi }&=&L_z\,,
\eea
\ees
Then, Eq.~\eqref{eq: separation1} becomes
\begin{widetext}
\bea
-\Sigma&=&-\frac{\left(r^2+a^2\right)^2-a^2 \sin ^2\theta (r^2A(u)+a^2)}{r^2A(u)+a^2}\hat{H}_{\rm eff}^2+2\frac{\widetilde{\omega}_{\rm fd}}{r^2A(u)+a^2}\hat{H}_{\rm eff} \hat{L}_z+ \Delta_r\hat{P_r}^2  \nonumber\\
&&+\hat{P_{\theta }}^2+\frac{1}{\left(r^2+a^2\right)^2-a^2 \sin ^2\theta  (r^2A(u)+a^2)}\left(\frac{(r^2+a^2 \cos ^2\theta)^2}{\sin ^2\theta }-\frac{{\widetilde{\omega}_{\rm fd}}^2}{r^2A(u)+a^2 }\right)\hat{L_{z}}^2\label{eq: separation2}\,,
\eea
where we have defined the reduced momenta $\hat{P}_r=P_r/\mu$, $\hat{P}_\theta=P_\theta/\mu$, and $\hat{L}_z=L_z/\mu$. It is convenient to rewrite this expression as
\bea
-\Sigma&=&-\frac{\left(a\hat{L_{z}}-\left(r^2+a^2\right)\hat{H}_{\rm eff}\right)^2}{r^2A(u)+a^2}+\frac{\left(\hat{L_{z}}-a\sin ^2\theta\hat{H}_{\rm eff}\right)^2}{\sin ^2\theta}+ \Delta_r\hat{P_r}^2+\hat{P_{\theta }}^2  \nonumber\\
&&+2\frac{\widetilde{\omega}_{\rm fd}+a r^2\left(A(u)-1\right)}{r^2A(u)+a^2}\hat{H}_{\rm eff}\hat{L_{z}}-\frac{{\widetilde{\omega}_{\rm fd}}^2-a^2r^4\left(A(u)-1\right)^2}{\left(r^2A(u)+a^2\right)\left(\left(r^2+a^2\right)^2-a^2 \sin ^2\theta  (r^2A(u)+a^2)\right)}\hat{L_{z}}^2\label{eq: separation3}\,,
\eea
and solving this equation by separation of variables gives
\be
\label{separation1}
\left(\frac{\hat{L_{z}}}{\sin \theta}\!-\!a\hat{H}_{\rm eff}\sin \theta\right)^2\!+\!\hat{P_{\theta }}^2\!+\!a^2\cos^2\theta\!=\!\frac{\left(a\hat{L_{z}}-\left(r^2+a^2\right)\hat{H}_{\rm eff}\right)^2}{r^2A(u)+a^2}-\Delta_r\hat{P_r}^2-r^2-2\frac{\widetilde{\omega}_{\rm fd}+a r^2\left(A(u)-1\right)}{r^2A(u)+a^2}\hat{H}_{\rm eff}\hat{L_{z}}+F(r,\theta)\hat{L_{z}}^2 \,,
\ee
where
\be
F(r,\theta)=\frac{{\widetilde{\omega}_{\rm fd}}^2-a^2r^4\left(A(u)-1\right)^2}{\left(r^2A(u)+a^2\right)\left(\left(r^2+a^2\right)^2-a^2 \sin ^2\theta  (r^2A(u)+a^2)\right)}\,.
\ee
\end{widetext}

In the test-particle limit $\nu \to 0$, $A(u)\to1-2u$, the $r$-$\theta$ coupling function $F(r,\theta)\to0$, and both sides must be equal to a new constant of the motion since their Poisson brackets with the Hamiltonian  are equal to zero, and thus
we can obtain the reduced Carter constant $\hat{\mathcal{K}}$ expressed in terms of ($\theta$, $\hat{P_{\theta }}$) like Eq.(49) of Ref.~\cite{carter68},
\be
\label{Kintheta}
\hat{\mathcal{K}}=\hat{P_{\theta }}^2+a^2\cos^2\theta+\left(\frac{\hat{L_{z}}}{\sin \theta}\!-\!a\sin \theta\hat{H}_{\rm eff}\right)^2 \,,
\ee
There is another definition of the Carter constant, $\hat{Q}\equiv\hat{\mathcal{K}}-(\hat{L_{z}}-a \hat{H_{\rm eff}})^2$, which vanishes for equatorial orbits($\theta=\pi/2$).

If we consider the cases of the inspirals with nonzero mass ratio, $F(r,\theta)$ should no longer be ignored, then the $r$ and $\theta$ motion cannot be decoupled anymore. Now, let us see if there is an approximation $F(r,\theta) \approx F(r)$. It is convenient to rewrite the function $F(r,\theta)$ as
\begin{widetext}
\be
F(r,\theta)=\frac{{\widetilde{\omega}_{\rm fd}}^2-a^2r^4\left(A(u)-1\right)^2}{\left(r^2A(u)+a^2\right)\left(r^2+a^2\right)^2}\times\frac{1}{1-\frac{a^2 \sin ^2\theta  \left(r^2A(u)+a^2\right)}{\left(r^2+a^2\right)^2}} \,.
\ee
\end{widetext}
It can be easily noticed that $\frac{a^2 \sin ^2\theta \left(r^2A(u)+a^2\right)}{\left(r^2+a^2\right)^2}$ is a 2PN correction and we abort it first, and thus the first-order Maclaurin series of $F(r,\theta)$ is obtained as
\be
\label{eq:Frthata1}
F(r,\theta)\simeq G(r)+\frac{a^2 \sin ^2\theta}{\left(r^2+a^2\right)^4} ({\widetilde{\omega}_{\rm fd}}^2-a^2r^4\left(A(u)-1\right)^2)\,,
\ee
where
\be
G(r)=\frac{{\widetilde{\omega}_{\rm fd}}^2-a^2r^4\left(A(u)-1\right)^2}{\left(r^2A(u)+a^2\right)\left(r^2+a^2\right)^2} \,, 
\ee
and through 3PN order~\cite{buonanno2000transition,damour2000determination}
\be
A(u)=1-2u+\nu\left( 2u^3+(\frac{94}{3}-\frac{41\pi^2}{32})u^4\right)\,,
\ee
Inserting it into Eq.~\eqref{eq:Frthata1} [truncated at $\mathcal{O}(\nu)$ term] leads to
\begin{widetext}
\be
F(r,\theta)\simeq G(r)+\frac{4 \,\nu \,a^4 M^4 \sin ^2\theta }{\left(r^2+a^2\right)^4}\left(2+\left(\frac{94}{3}-\frac{41 \pi ^2}{32}\right)u+\omega_1^{\rm fd}+ \omega_2^{\rm fd}\frac{a^2}{M^2}\right) \,, \ee
where the second term as a function of both $r$ and $\theta$ is just 8PN with a mass ratio and can be safely ignored. We replace $F(r,\theta)$ by G(r) to decouple $r$ and $\theta$ so that Eq.~\eqref{separation1} can be separated. As shown in Fig.~\ref{fig:1}, the error between $F(r,\theta)$ and $G(r)$ mainly depends on the value of $r$ and is maximal when $\theta = \pi/2$ (the equatorial plane). One can see that even very close to the horizon  ($\xi \rightarrow 0$ for the orange line), the error is still one order smaller than the mass ratio. Otherwise, the term we aborted is two or more orders less than the mass ratio. We believe that replacing $F(r,\theta)$ with $G(r)$  retains enough accuracy for EMRIs. 

\begin{figure}[!h]
\begin{center}

\includegraphics[width=0.8\columnwidth]{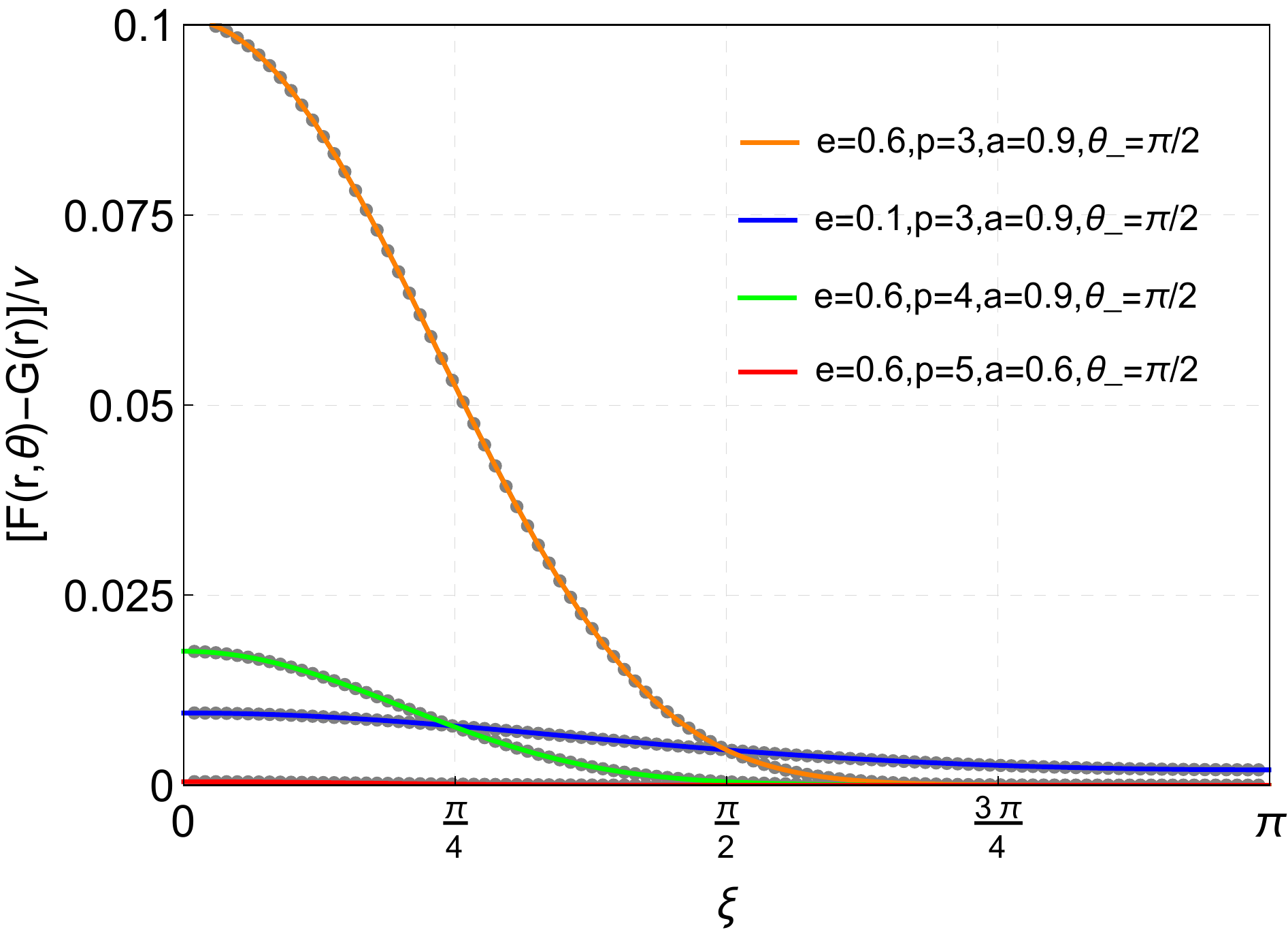}
\caption{\emph{Error of the approximation of the coupling function $F(r,\theta)$.} The solid line and points represent $\nu =10^{-4}$ and $10^{-6}$ respectively. It is a very close approximation to substitute the r function $G(r)$ for $F(r,\theta)$. Even if we chose the extreme orbital parameters near LSO(for the orange line, $p_s=2.96M$), the error is less than $0.1\nu$.}
\label{fig:1}
\end{center}
\end{figure}

Now we can obtain the approximate reduced Carter constant $\hat{\mathcal{K}}$ expressed in terms of ($\theta$, $\hat{P_{\theta }}$) same as Eq.~\eqref{Kintheta} and the other expression in terms of ($r$, $\hat{P_r}$)
\be
\hat{\mathcal{K}}=\frac{\Big[a\hat{L_{z}}\!-\!\left(r^2+a^2\right)\hat{H}_{\rm eff}\Big]^2}{r^2A(u)+a^2}-\Delta_r\hat{P_r}^2-r^2-2\frac{\widetilde{\omega}_{\rm fd}+a r^2\left(A(u)-1\right)}{r^2A(u)+a^2}\hat{H}_{\rm eff}\hat{L_{z}}\!\!+G(r)\hat{L_{z}}^2 \,,
\ee
\end{widetext}

By using these constants of motion, the angular momentum $\hat{P_{\theta}}$ and $\hat{P_r}$ can be expressed as
\begin{widetext}
\bes
\bea
&\hat{P_{\theta}}^2=\hat{Q}-\cos ^2\theta  \bigg[a^2 \left(1-\hat{H}_{\rm eff}^2\right)+\cfrac{\hat{L_{z}}^2}{\sin^2 \theta}\bigg]\,\label{eq:ptheta},\\
&\hat{P_r}^2=\cfrac{\Big[a\hat{L_{z}}\!-\!(r^2\!+\!a^2)\hat{H_{\rm eff}}\Big]^2\!-\!\left( r^2 A(u)\!+\!a^2\right)\bigg[r^2\!+\!\hat{\mathcal{K}}\!+\!2\cfrac{\widetilde{\omega}_{\rm fd}+a r^2\left(A(u)-1\right)}{r^2A(u)+a^2}\hat{H}_{\rm eff}\hat{L_{z}}\!-\!G(r)\hat{L_{z}}^2\bigg]}{\left( r^2 A(u)\!+\!a^2\right)^2 D^{-1}(u) }\,\label{eq:pr}.
\eea
\ees
\end{widetext}
In Eq.~\eqref{eq:ptheta}, if $\hat{P}_\theta = 0$, the polar motion is at the turning points $\theta = \theta_{\rm min}$ or $\pi-\theta_{\rm min}$. Then we get the relation between the semi-Carter ``constant" and the orbital inclination,
\bea
\hat{Q} = \cos ^2\theta_{\rm min}  \bigg[a^2 \left(1-\hat{H}_{\rm eff}^2\right)+\cfrac{\hat{L_{z}}^2}{\sin^2 \theta_{\rm min}}\bigg]\,.\label{eq:carter}
\eea
This semi-Carter constant has the same form as in the test-particle case, but with mass-ratio corrections hidden in $H_{\rm eff}$ and $L_z$, which will be given explicitly later.

There is another straightforward way to get $\boldmath{P}$ by numerically solving the Hamiltonian equations (\ref{eq:hamilton}). We then compare the analytical equation  Eq.~\eqref{eq:ptheta} and the numerical results (also including the spin terms like as $H_s$) and show the results in Table~\ref{ananlytical numerical}. We find that our analytical
approximations are very close to the numerical results, even for a mass ratio as large as 0.01, which proves that our approximation works well for EMRIs or even intermediatemass-ratio inspirals. 

\begin{widetext}
\begin{table}[!h]
\begin{center}
\caption{Comparison of the maximum $\hat{P_{\theta }}$ ($\hat{P_{\theta }}$ at $\theta=\pi/2$) between the analytical formalism and numerical integration.}
\begin{tabular}{|c|c|c|}
\hline $\nu=0.01$&analytical max$\hat{P_{\theta }}$&numerical max$\hat{P_{\theta }}$\\ \hline
$p=8,e=1/3,a=0.99,\theta_{\rm min}=\pi/4$&$2.2909944716$&$2.2909945165$\\ \hline
$p=4,e=1/3,a=0.99,\theta_{\rm min}=\pi/4$&$1.8126129568$&$1.8126323323$\\ \hline
$p=8,e=1/3,a=0.1,\theta_{\rm min}=\pi/4$&$2.5258065572$&$2.5258065544$\\ \hline
$p=8,e=1/3,a=0.99,\theta_{\rm min}=\pi/10$&$3.2445223467$&$3.2445223751$\\ \hline
\end{tabular}\label{ananlytical numerical}
\end{center}
\end{table}
\end{widetext}

\subsection{Reparametrization of the energy and angular
momentum}
\label{sec:constants}

%
Solving Eqs.~\eqref{eq:superhamiltonian} and~\eqref{eq:pt} for ${H}_{\rm eff}$ yields the effective Hamiltonian  associated with the deformed Kerr metric 
\begin{equation}\label{eq:Hns}
{H}_{\rm eff} = \frac{g^{t \phi}}{g^{tt}} P_\phi + \frac{1}{\sqrt{-g^{tt}}} \sqrt{\mu^2 + \Bigg[g^{\phi\phi}-\frac{(g^{t \phi})^2}{g^{tt}} \Bigg]P_\phi^2+g^{rr}P_r^2+g^{\theta\theta }P^2_\theta }\,,
\end{equation}
The energy of the system is given by
\be
E=H_{\rm EOB},
\ee
which implies the relation
\be
\hat{H}_{\rm eff}(E)=1+\frac{1}{2\nu}\left(\frac{E^2}{M^2}-1\right). \label{eq:HeffofE}
\ee

\begin{figure}[!h]
\begin{center}
\includegraphics[width=0.8\columnwidth]{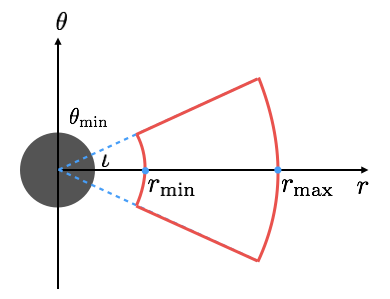}
\caption{\emph{Diagram of 3D motion of a CO around a Kerr black hole.} The area encircled by the solid red line is the section of trajectories on the $r$-$\theta$ plane. Here, $r_{\rm max/min}$ is just $r_{1/2}$ in Eq. (\ref{eq:r12}).}
\label{fig:orbit3d}
\end{center}
\end{figure}

The constants of motion and the dynamical equations in the last subsection can be written in terms of the geometrized orbital elements: the semilatus rectum $p$,  eccentricity $e$, and orbital inclination $\iota$. This will make the description of the system more intuitive. For an eccentric orbit, there exist apastron and periastron points which can be expressed as
\be
r_{1}=\frac{p M}{1-e}, \ \ \ \ \ r_2=\frac{p M}{1+e}, \label{eq:r12}
\ee
where\;$r_1,r_2$\;are the turning points of the radial motion. $\theta_{\rm min}$ and $\pi-\theta_{\rm min}$ is the turning points of polar motion, and $\iota \equiv \pi/2-\theta_{\rm min}$ defines the so-called orbital inclination (see Fig.~\ref{fig:orbit3d}). These turning points are computed by solving the radial and polar equations of motion for\;$\dot{r}=0, \dot{\theta}=0$. By setting the radial and polar equations of motion equals to zero with $P_r=P_\theta=0$, evaluating Eq.~\eqref{eq:HeffofE} at $(r_1,\theta_{\rm min})$ and $(r_2,\theta_{\rm min})$ leads to

\begin{widetext}
\bes
\label{eq:psofep}\begin{align}
\hat{L}_z^2\!&=\!\frac{\left(a_1\!-\!a_2\right)^2\left(b_1^2\!+\!b_2^2\right)\!-\!(b_1^2\!-\!b_2^2)(b_1^2c_1\!-\!b_2^2c_2)\!-\!2 (a_1\!-\!a_2)b_1 b_2\sqrt{\left(a_1\!-\!a_2\right)^2\!-\!\left(b_1^2\!-\!b_2^2\right)\left(c_1\!-\!c_2\right)}}{\left[(a_1\!-\!a_2)^2\!-\!\left(b_1^2 c_1\!-\!b_2^2 c_2\right)\right]^2}  \\
\frac{ E^2}{M^2}\!&=\!1+2\nu\left(a_1 \hat{L}_z\!+\!\sqrt{\frac{c_1 \hat{L}_z^2\!+\!1}{b_1} } -1\right)\qquad
\end{align}
\ees
\end{widetext}

where the coefficients are
\bes
\label{eq:abc}
\bea
a_1&&=\frac{\widetilde{\omega}_{\rm fd1} }{\Lambda _{t1}}\,,\\
a_2&&=\frac{\widetilde{\omega}_{\rm fd2} }{\Lambda _{t2}}\,,\\
b_1&&=\sqrt{\frac{\Sigma _1 \Delta _{t1}}{\Lambda _{t1}}}\,,\\
b_2&&=\sqrt{\frac{\Sigma _2 \Delta _{t2}}{\Lambda _{t2}}}\,,\\
c_1&&=\frac{\Sigma _1}{\left(1-\cos ^2\theta_{\rm min}\right)\Lambda _{t1}}\,,\\
c_2&&=\frac{\Sigma _2}{\left(1-\cos ^2\theta_{\rm min}\right)\Lambda _{t2}},
\eea
\ees
in which the subscripts 1 and 2 mean that the function is to be evaluated at $(r_1,\theta_{\rm min})$ and $(r_2,\theta_{\rm min})$,
\bes
\bea
\Sigma _1&&=a^2 \cos ^2\theta_{\rm min}+ r_1^2\,,\\
\Sigma _2&&=a^2 \cos ^2\theta_{\rm min}+ r_2^2\,,\\
\Delta _{t1}&&=a^2+A(r_1) r_1^2\,,\\
\Delta _{t2}&&=a^2+A(r_2) r_2^2\,,\\
\Lambda _{t1}&&=\left(a^2+r_1^2\right)^2-a^2 \sin ^2\theta_{\rm min} \Delta _{t1}\,,\\
\Lambda _{t2}&&=\left(a^2+r_2^2\right)^2-a^2 \sin ^2\theta_{\rm min}  \Delta _{t2}\,,\\
\widetilde{\omega}_{\rm fd1}&&=\frac{a^3 \nu  \omega _2}{r_1}+\frac{a \nu  \omega _1}{r_1}+2 a r_1\,,\\
\widetilde{\omega}_{\rm fd2}&&=\frac{a^3 \nu  \omega _2}{r_2}+\frac{a \nu  \omega _1}{r_2}+2 a r_2.
\eea
\ees

The above formalism for a Kerr black hole is much more complicated than the Schwarzschild ones in Ref.~\cite{hinderer2017foundations}. Obviously, for the test-particle limit $\nu \to 0$, the above results will go back to the geodesic motion of a test particle in Kerr spacetime.

\subsection{Determination of the fundamental frequencies}
\label{sec:frequencies}

Since in our approximation the equations of motion in the deformed-Kerr spacetime are separable in the coordinates $r$, $\theta$, and $\phi$, the action variables can be calculated from cyclic integrals over the spatial conjugate momenta in the Boyer-Lindquist coordinate representation:
\bea
  \label{eq:action_var_r}
  J_{r} && = \frac{1}{2\pi}\oint P_{r}\,\rmd r\,, \\
  \label{eq:action_var_theta}
  J_{\theta}&& = \frac{1}{2\pi}\oint
  P_{\theta}\,\rmd \theta\,, \\
  \label{eq:action_var_phi}
  J_{\phi}  && = \frac{1}{2\pi}\oint P_{\phi}\,\rmd \phi = L_{z}\,.
\eea
The standard procedure of determining fundamental frequencies is to find the explicit form of the Hamiltonian in the action-angle
representation, $H^{(\mathrm{aa})}$, and calculate the frequencies from the partial derivatives with respect to the radial and polar action variables
$J_{k}$~\cite{goldstein1980classical}:
\be
\label{eq:frequency}
  \mu\omega_{k} = \frac{\partial H}{\partial J_{k}}^{(\mathrm{aa})}\,.
\ee
Unfortunately, Eqs.~\eqref{eq:action_var_r} and~\eqref{eq:action_var_theta} cannot be solved analytically, and so they do not admit an explicit inversion. However, according to the Schmidt method~\cite{Schmidt2002Celestial}, Eq.~\eqref{eq:frequency} can be solved even without knowing the functional form of the Hamiltonian $H^{(\mathrm{aa})}$ if the theorem on implicit functions is employed.

Let
$P_{\beta}^{(\mathrm{aa})}=f_{\beta}^{(\mathrm{aa})}(-\mu^{2}/2,H_{\rm eff},L_{z},Q)$
be the momenta given by $P_{0}^{(\mathrm{aa})}=p_{t}=-H_{\rm eff}$ and
$P_{k}^{(\mathrm{aa})}=J_{k}$. If we
denote the Jacobian matrix of $f$ by $\mathsfsl{D}f$, then, by the
theorem on implicit functions,
$\mathsfsl{D}f\cdot\mathsfsl{D}(f^{-1})=
\mathsfsl{D}f\cdot(\mathsfsl{D}f)^{-1}=\mathsfsl{I}$, provided that
$f$ is non-zero and the Jacobian does not vanish~\cite{1979Analysis}. As
$-\mu^2/2$ is the invariant value of the Hamiltonian, we can
substitute $-\mu^2/2=g^{\alpha \beta}P_\alpha P_\beta/2=H^{(\mathrm{aa})}(-H_{\rm eff},J_{k})$. In addition, two rows of the Jacobian matrix are trivial due to the identities
$P_{0}^{(\mathrm{aa})}=-H_{\rm eff}$ and $J_{\phi}=L_{z}$. For simplicity, we use the symbol $H$ to denote the Hamiltonian $H^{(\mathrm{aa})}$ below. Thus, the equation
$\mathsfsl{D}f\cdot\mathsfsl{D}(f^{-1})=\mathsfsl{I}$ reads
\begin{equation}
  \left(\begin{array}{llll}
    0 & -1 & 0 & 0 \\
    \frac{\partial J_{r}}{\partial H} & 
    \frac{\partial J_{r}}{\partial H_{\rm eff}} &
    \frac{\partial J_{r}}{\partial L_{z}} & 
    \frac{\partial J_{r}}{\partial Q} \\
    \frac{\partial J_{\theta}}{\partial H} & 
    \frac{\partial J_{\theta}}{\partial H_{\rm eff}} &
    \frac{\partial J_{\theta}}{\partial L_{z}} & 
    \frac{\partial J_{\theta}}{\partial Q} \\
    0 & 0 & 1 & 0
  \end{array}\right)\cdot
  \left(\begin{array}{llll}
    -\frac{\partial H}{\partial H_{\rm eff}} &
    \frac{\partial H}{\partial J_{r}} &
    \frac{\partial H}{\partial J_{\theta}} &
    \frac{\partial H}{\partial J_{\phi}} \\
    -1 & 0 & 0 & 0 \\
    0 & 0 & 0 & 1 \\
    -\frac{\partial Q}{\partial H_{\rm eff}} &
    \frac{\partial Q}{\partial J_{r}} &
    \frac{\partial Q}{\partial J_{\theta}} &
    \frac{\partial Q}{\partial J_{\phi}} \\
   \end{array}\right) = \mathsfsl{I}.
\end{equation}
The above matrix equation is valid if and only if the orbit is nonequatorial~\cite{Schmidt2002Celestial}. We have discussed the condition of equatorial orbits in our previous work~\cite{zc2020}. Then, it can be split into four nontrivial sets of linear equations in the eight unknowns $-\frac{\partial H}{\partial
  H_{\rm eff}}$, $\frac{\partial H}{\partial J_{k}}$, $-\frac{\partial
  Q}{\partial H_{\rm eff}}$, and $\frac{\partial Q}{\partial J_{k}}$:
\begin{eqnarray}
  -&&\mathsfsl{A}\cdot\frac{\partial}{\partial H_{\rm eff}}
  \left(\begin{array}{llll}  H \\ Q \end{array}\right) =
  \left(\begin{array}{llll} 
     2 W(r_{1},r_{2}) \\
    2 a^{2} H_{\rm eff} U(\theta_{\rm min},\pi/2)
  \end{array}\right) \\
  \label{eq:syst_act_r}
  &&\mathsfsl{A}\cdot\frac{\partial}{\partial J_{r}}
  \left(\begin{array}{llll}  H \\ Q \end{array}\right) =
  \left(\begin{array}{llll} 2\pi \\ 0 \end{array}\right), \\
  &&\mathsfsl{A}\cdot\frac{\partial}{\partial J_{\theta}}
  \left(\begin{array}{llll}  H \\ Q \end{array}\right) =
  \left(\begin{array}{llll} 0 \\ \pi \end{array}\right), \\
  &&\mathsfsl{A}\cdot\frac{\partial}{\partial J_{\phi}}
  \left(\begin{array}{llll}  H \\ Q \end{array}\right) =
  \left(\begin{array}{llll}
   2 Z(r_{1},r_{2})  \\
    2 L_{z} V(\theta_{\rm min},\pi/2)
  \end{array}\right),
\end{eqnarray}
with the coefficient matrix
\begin{equation}
  \mathsfsl{A} = \left(\begin{array}{llll}
   2\pi \frac{\partial J_{r}}{\partial H} & 
   2\pi \frac{\partial J_{r}}{\partial Q} \\
   \pi \frac{\partial J_{\theta}}{\partial H} & 
   \pi \frac{\partial J_{\theta}}{\partial Q} 
  \end{array}\right)=
  \left(\begin{array}{llll} 
    2Y(r_{1},r_{2}) & 
    -X(r_{1},r_{2}) \\ 
    2 a^{2}U(\theta_{\rm min},\pi/2)&
    T(\theta_{\rm min},\pi/2)
  \end{array}\right),
\end{equation}

where $r_{1}$ and $r_{2}$ are the turning points of radial motion, and $X(r_{1},r_{2})$, $Y(r_{1},r_{2})$,$Z(r_{1},r_{2})$ and $W(r_{1},r_{2})$ are radial integrals defined by
\bes
\bea
  X(r_{1},r_{2}) & = &\int_{r_{1}}^{r_{2}}\frac{\rmd r}{\Delta_r P_r}, \\
  Y(r_{1},r_{2}) & = &\int_{r_{1}}^{r_{2}}\frac{r^{2}\,\rmd r}{\Delta_r P_r}, \\
  Z(r_{1},r_{2}) & = -&\int_{r_{1}}^{r_{2}}
  \frac{\partial P_r}{\partial L_z}\,\rmd r, \\
  W(r_{1},r_{2})& = &
  \int_{r_{1}}^{r_{2}}\frac{\partial P_r}{\partial H_{\rm eff}}\,\rmd r,
\eea
\ees
$T(\theta_{\rm min},\pi/2)$, $U(\theta_{\rm min},\pi/2)$, and $V(\theta_{\rm min},\pi/2)$ are polar integrals defined by
\bes
\bea
 T(\theta_{\rm min},\pi/2)& = &\int_{\theta_{\rm min}}^{\pi/2}\frac{\rmd \theta}{P_\theta}, \\
 U(\theta_{\rm min},\pi/2)& = &\int_{\theta_{\rm min}}^{\pi/2}\frac{\cos^{2}\theta}{P_\theta}\,\rmd \theta, \\
 V(\theta_{\rm min},\pi/2)& = &\int_{\theta_{\rm min}}^{\pi/2}\frac{\cot^{2}\theta}{P_\theta}\,\rmd \theta   .
\eea
\ees
The above radial functions $X,Y,Z,W$ and polar functions $T,U,V$ are not proper integrals because the integrated functions are divergent at the turning points $r_1,r_2$ and $\theta_{\rm min},\pi-\theta_{\rm min}$. Thus, we define $\xi$ by the equation $r=pM/(1+e\cos\xi)$, where $p$ is called the semi-latus rectum and $e$ is the eccentricity of the orbit, and we define $\chi$ with $\cos^2\theta=\cos^2\theta_{\rm min}\cos^2\chi$. As $\xi$ varies from $0$ to $2\pi$ as $r$ goes through a complete cycle, $\chi$ varies from $0$ to $2\pi$ as $\theta$ oscillates through its full range of motion. Then transform $X,Y,X,W$ and $T,U,V$ into well-behaved integrals
\bes
\begin{eqnarray}
\label{eq:rad_integr_x}
  X & = &\int_{0}^{\pi}\frac{ e p\sin \xi}{(1+e \cos \xi)^2 }\frac{\rmd \xi}{\Delta_r P_r}, \\
  \label{eq:rad_integr_y}
  Y & = &\int_{0}^{\pi}\frac{ e p\sin \xi}{(1+e \cos \xi)^2 }\frac{r^{2}\,\rmd \xi}{\Delta_r P_r}, \\  
  \label{eq:rad_integr_z}
  Z & = -&\int_{0}^{\pi}\frac{ e p\sin \xi}{(1+e \cos \xi)^2 }
  \frac{\partial P_r}{\partial L_z}\,\rmd \xi, \\
    \label{eq:rad_integr_w}
  W & = &
  \int_{0}^{\pi}\frac{ e p\sin \xi}{(1+e \cos \xi)^2 }\frac{\partial P_r}{\partial H_{\rm eff}}\,\rmd \xi, 
\end{eqnarray}
\ees
which transforming Eq.~\eqref{eq:ptheta} to
\be
\beta^2(z^2-z_+^2)(z^2-z_-^2)=P_{\theta}^2(1-z^2)  \label{eq:pthetaz}\,, 
\ee
where $\beta^{2}=a^{2}(\mu^2-{H_{\rm eff}}^{2})$, $z_{\pm}^{2}$ are the two roots of the equation $P_\theta=0$ when substituting $\cos\theta=z$ in $P_\theta$ and given by 
\bea
z_-&=&\cos\theta_{\rm min}\,, \\
z_+&=&\cfrac{L_z^2+Q+\beta+\sqrt{(L_z^2+Q+\beta)^2-4\beta Q}}{2\beta}\,,
\eea 
by this way, Schmidt derives the following expressions~\cite{Schmidt2002Celestial} for the polar integrals
\bes
\begin{eqnarray}
 T & = &
  \frac{1}{\beta z_{+}}K(k)\,,  \\
  \label{eq:angl_intgr2}
  U & = &
  \frac{z_{+}}{\beta}[K(k)-E(k)]\,,  \\
  \label{eq:angl_intgr3}
 V & = &
  \frac{z_{+}}{\beta}[\Pi(z_{-}^{2},k)-K(k)]\,, 
\end{eqnarray}
\ees
where $k=z_{-}^{2}/z_{+}^{2}$, $K(k)$, $E(k)$ and
$\Pi(z_{-}^{2},k)$ are, respectively, the complete elliptical
integrals of the first, second and third kind,
\bes
\bea
  \label{eq:ell_integr_k}
  K(k) & =&\int_{0}^{\pi/2}\!\frac{\rmd \chi}{\sqrt{1-k\sin^{2}\chi}}, \\
  \label{eq:ell_integr_e}
  E(k) & =&\int_{0}^{\pi/2}\!\sqrt{1-k\sin^{2}\chi}\,\rmd \chi, \\
  \label{eq:ell_integr_pi}
  \Pi({z_{-}^{2},k}) & =&
  \int_{0}^{\pi/2}\!\frac{\rmd\chi}{\left(1-z_{-}^{2}\sin^{2}\chi\right)
  \sqrt{1-k\sin^{2}\chi}},
\eea
\ees

The solutions of the above systems of equations for $-\frac{\partial
  H}{\partial H_{\rm eff}}$ and $\frac{\partial H}{\partial J_{k}}$ are given by
\begin{eqnarray}
  \label{eq:h_derv_p0}
  -&&\frac{\partial H}{\partial H_{\rm eff}} =
  \frac{K(k)W+ 
        a^{2}z_{+}^{2}E\left[K(k)-E(k)\right]X
        }{K(k)Y +
          a^{2}z_{+}^{2}\left[K(k)-E(k)\right]X}, \\
  \label{eq:h_derv_jr}
  &&\frac{\partial H}{\partial J_{r}} =
  \frac{\pi K(k)}{K(k)Y + a^{2}z_{+}^{2}\left[K(k)-E(k)\right]
                  X}, \\
  \label{eq:h_derv_jtheta}
  &&\frac{\partial H}{\partial J_{\theta}} =
  \frac{\pi\beta z_{+}X}{2\{K(k)Y +
    a^{2}z_{+}^{2}\left[K(k)-E(k)\right]X\}}, \\
  \label{eq:h_derv_jphi}
  &&\frac{\partial H}{\partial J_{\phi}} =
  \frac{K(k)Z + L_{z}[\Pi(z_{-}^{2},k)-K(k)]X
        }{K(k)Y + a^{2}z_{+}^{2}[K(k)-E(k)]X}.
\end{eqnarray}

From Eqs.~\eqref{eq:frequency} and~\eqref{eq:h_derv_p0}-\eqref{eq:h_derv_jphi}, the coordinate-time frequencies $\omega _r,\omega _{\theta}$ and $\omega _{\phi}$ are obtained
\bea
\omega _r &&=\frac{\pi K(k)}{K(k)W+ 
        a^{2}z_{+}^{2}E\left[K(k)-E(k)\right]X}, \\
\omega _{\theta} &&=\frac{\pi\beta z_{+}X}{2\{K(k)W+ 
        a^{2}z_{+}^{2}E\left[K(k)-E(k)\right]X\}}, \\
\omega _{\phi} &&=  \frac{K(k)Z + L_{z}[\Pi(z_{-}^{2},k)-K(k)]X
        }{K(k)W+a^{2}z_{+}^{2}E\left[K(k)-E(k)\right]X}.
\eea

\begin{figure}[!h]
\begin{center}
\includegraphics[width=0.5\columnwidth]{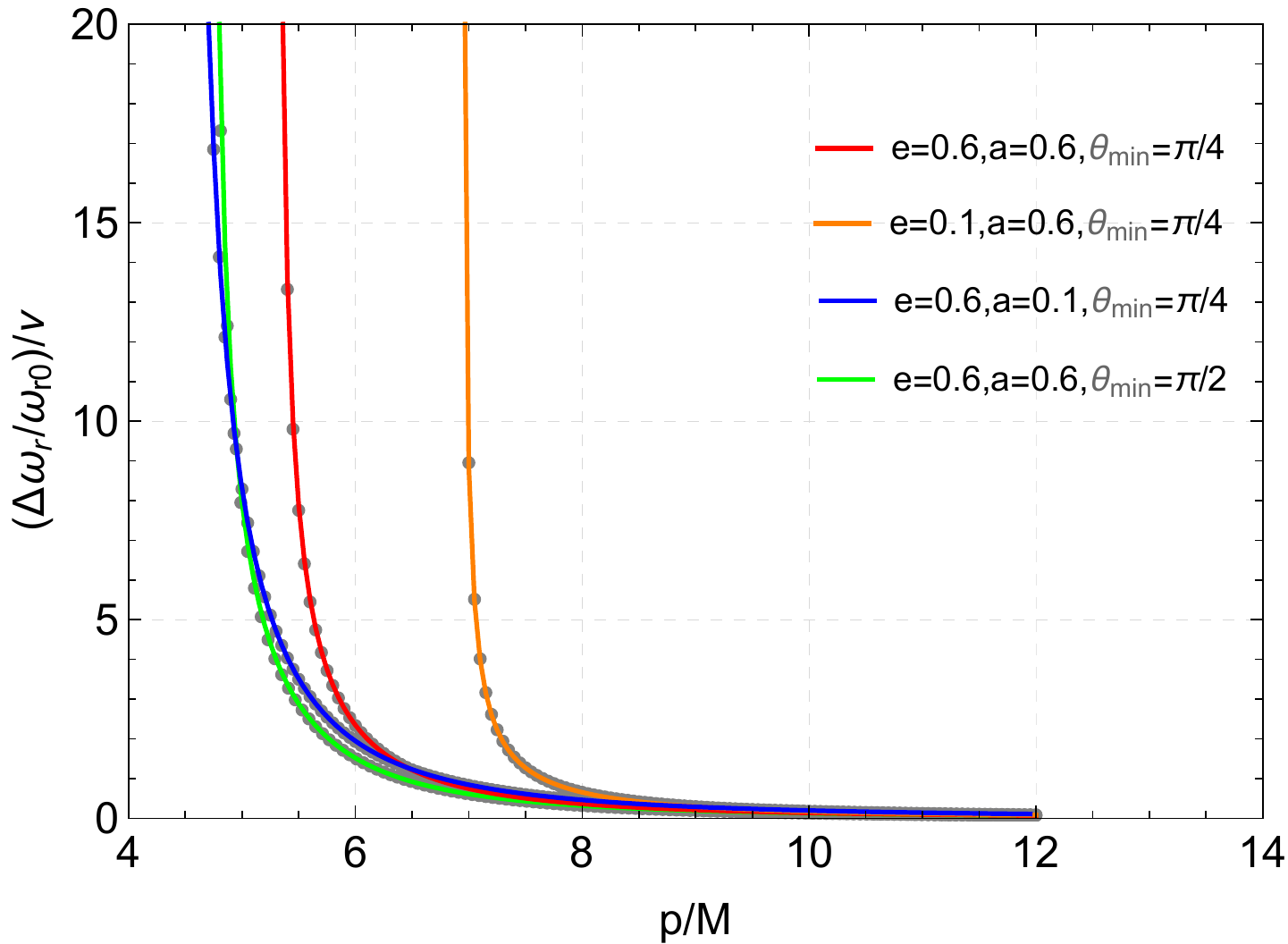}
\includegraphics[width=0.5\columnwidth]{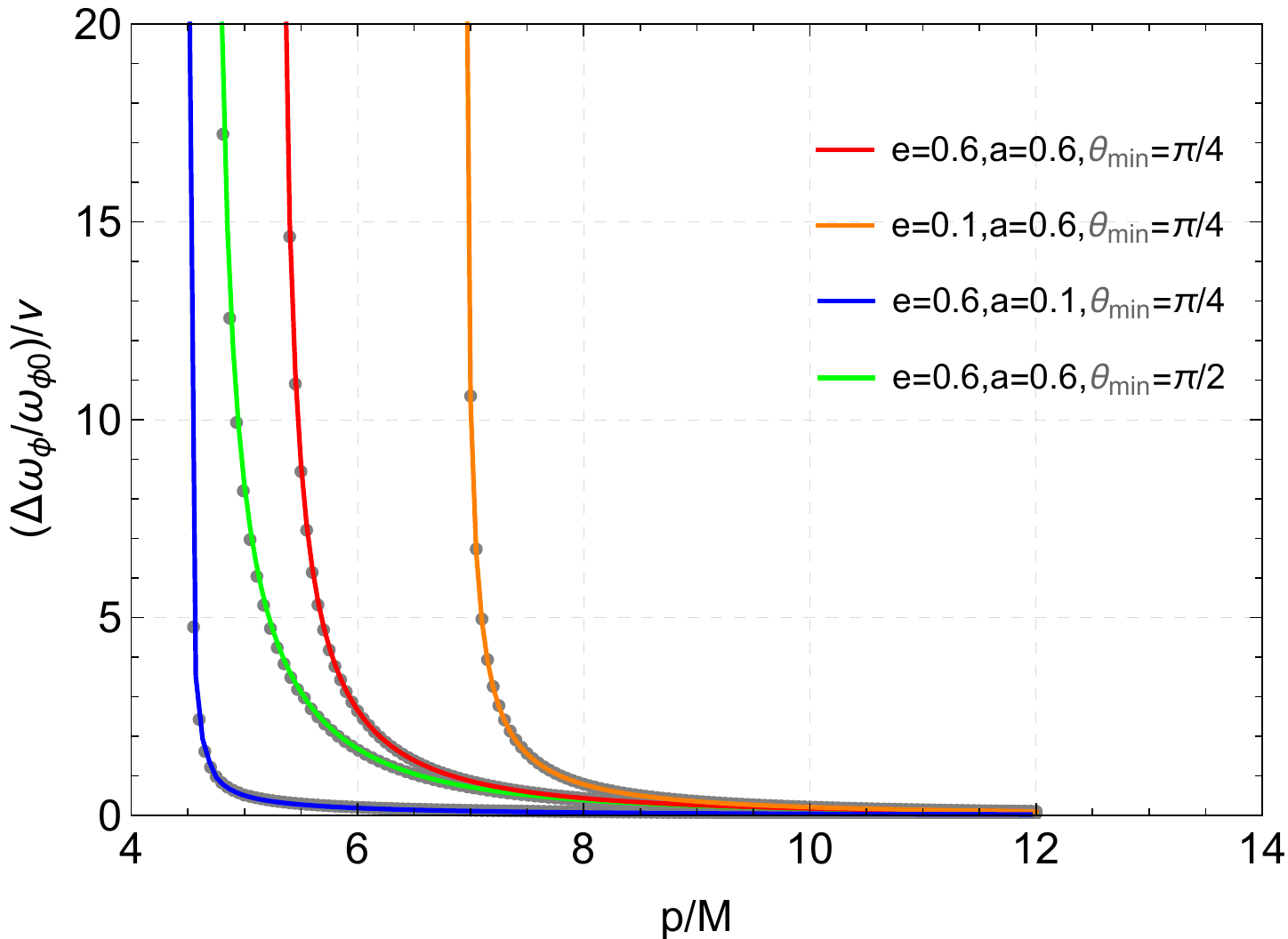}
\includegraphics[width=0.5\columnwidth]{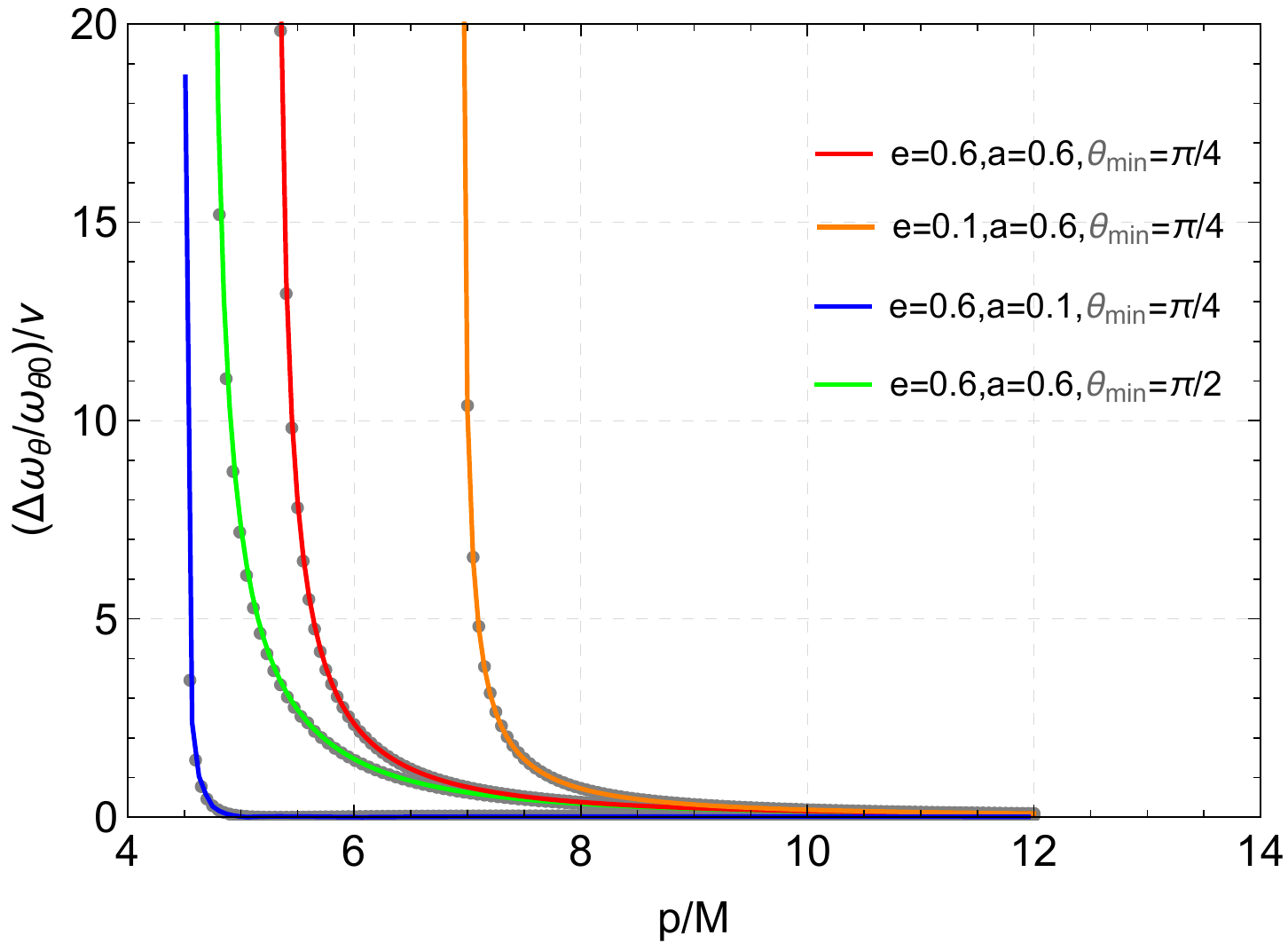}
\caption{\emph{Frequency shifts $\frac{\Delta\omega}{\nu\omega_{0}}$ vs semilatus rectum $p$ in the cases of various $a,~\nu,~ e,~ \theta_{\rm min}$.} The solid line and points represent $\nu =10^{-4}$ and $10^{-6}$ respectively.}
\label{fig:omega_p}
\end{center}
\end{figure}
\comment{
\begin{figure}[!h]
\begin{center}
\includegraphics[width=0.5\columnwidth]{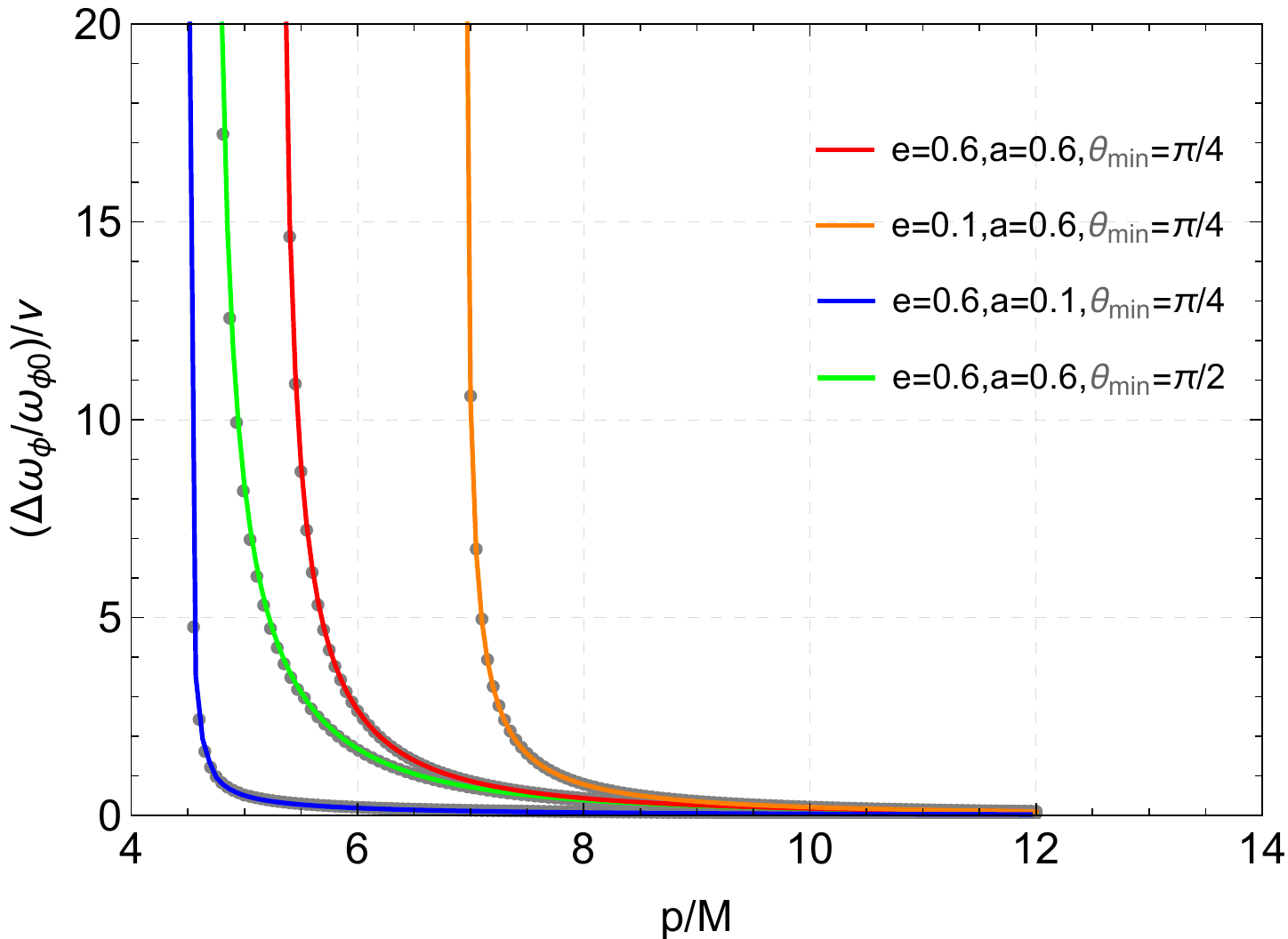}
\caption{\emph{Frequency shifts $\frac{\Delta\omega_{\phi}}{\nu\omega_{\phi0}}$ vs semilatus rectum $p$ in the cases of various $a,~\nu,~ e,~ \theta_{\rm min}$.} The solid line and points represent $\nu =10^{-4}$ and $10^{-6}$ respectively.}
\label{fig:3}
\end{center}
\end{figure}
\begin{figure}[!h]
\begin{center}
\includegraphics[width=0.5\columnwidth]{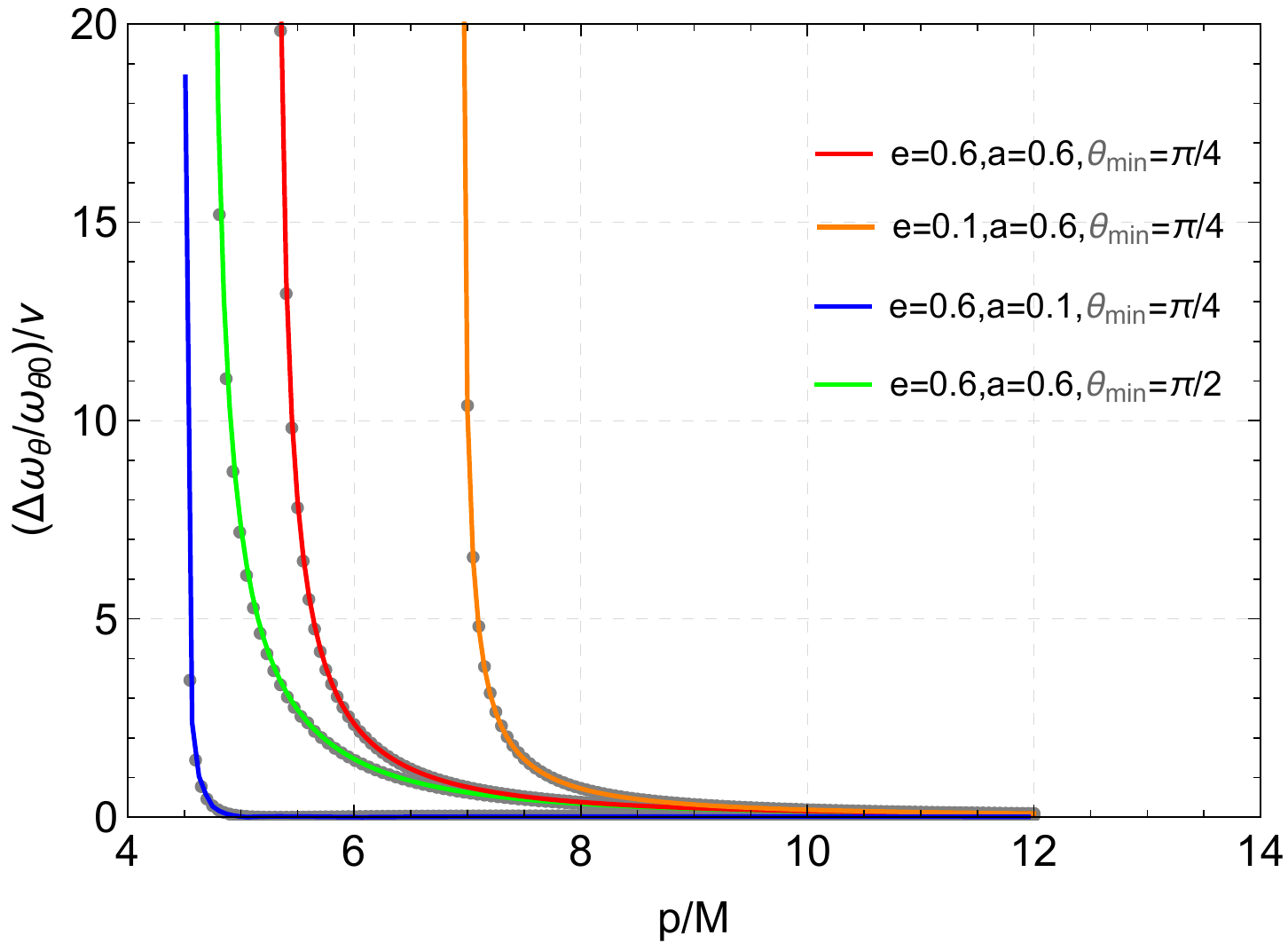}
\caption{\emph{Frequency shifts $\frac{\Delta\omega_{\theta}}{\nu\omega_{\theta0}}$ vs semilatus rectum $p$ in the cases of various $a,~\nu,~ e,~ \theta_{\rm min}$.} The solid line and points represent $\nu =10^{-4}$ and $10^{-6}$ respectively.}
\label{fig:4}
\end{center}
\end{figure}
}

\begin{figure}[!h]
\begin{center}
\includegraphics[width=0.5\columnwidth]{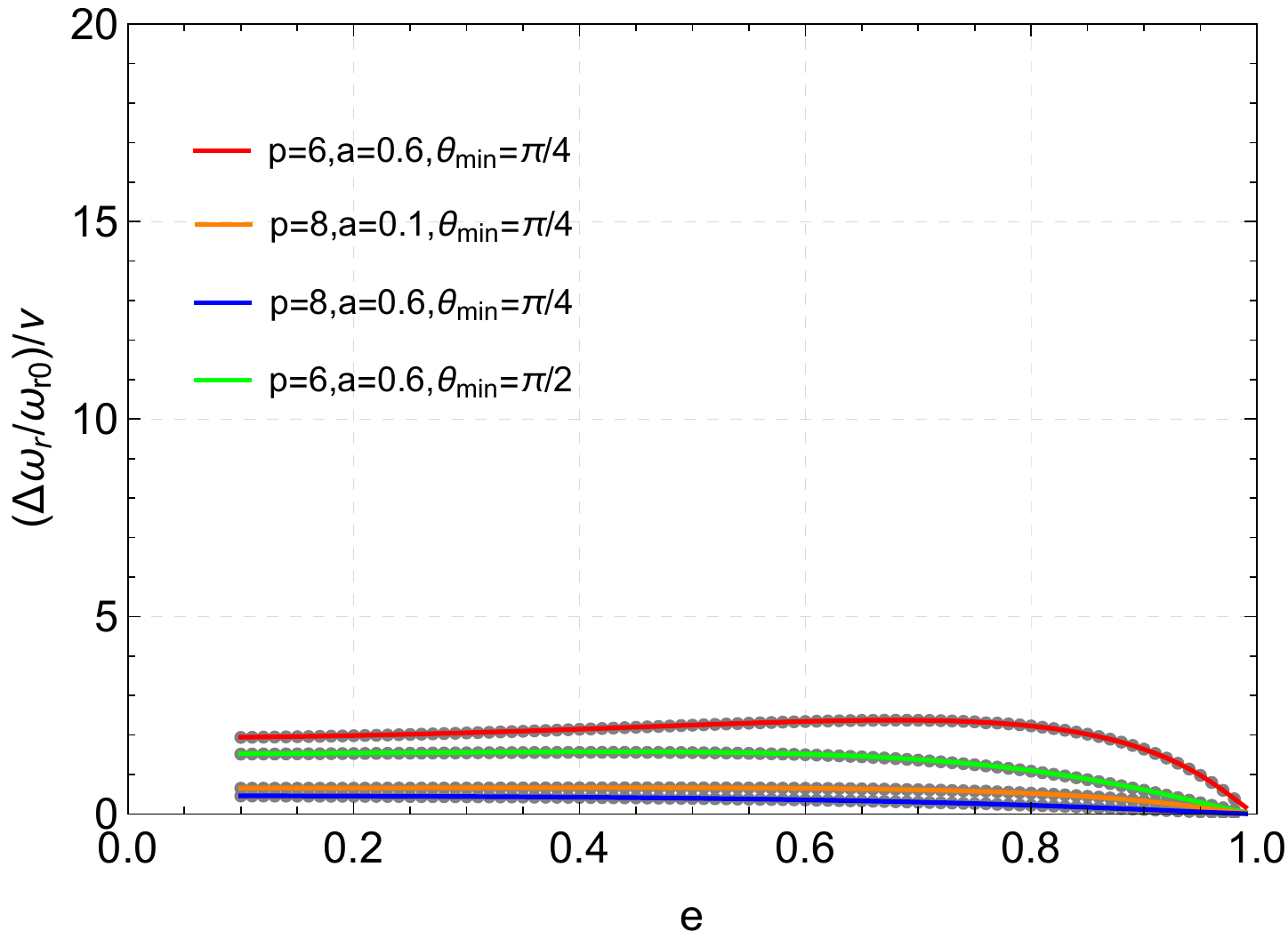}
\includegraphics[width=0.5\columnwidth]{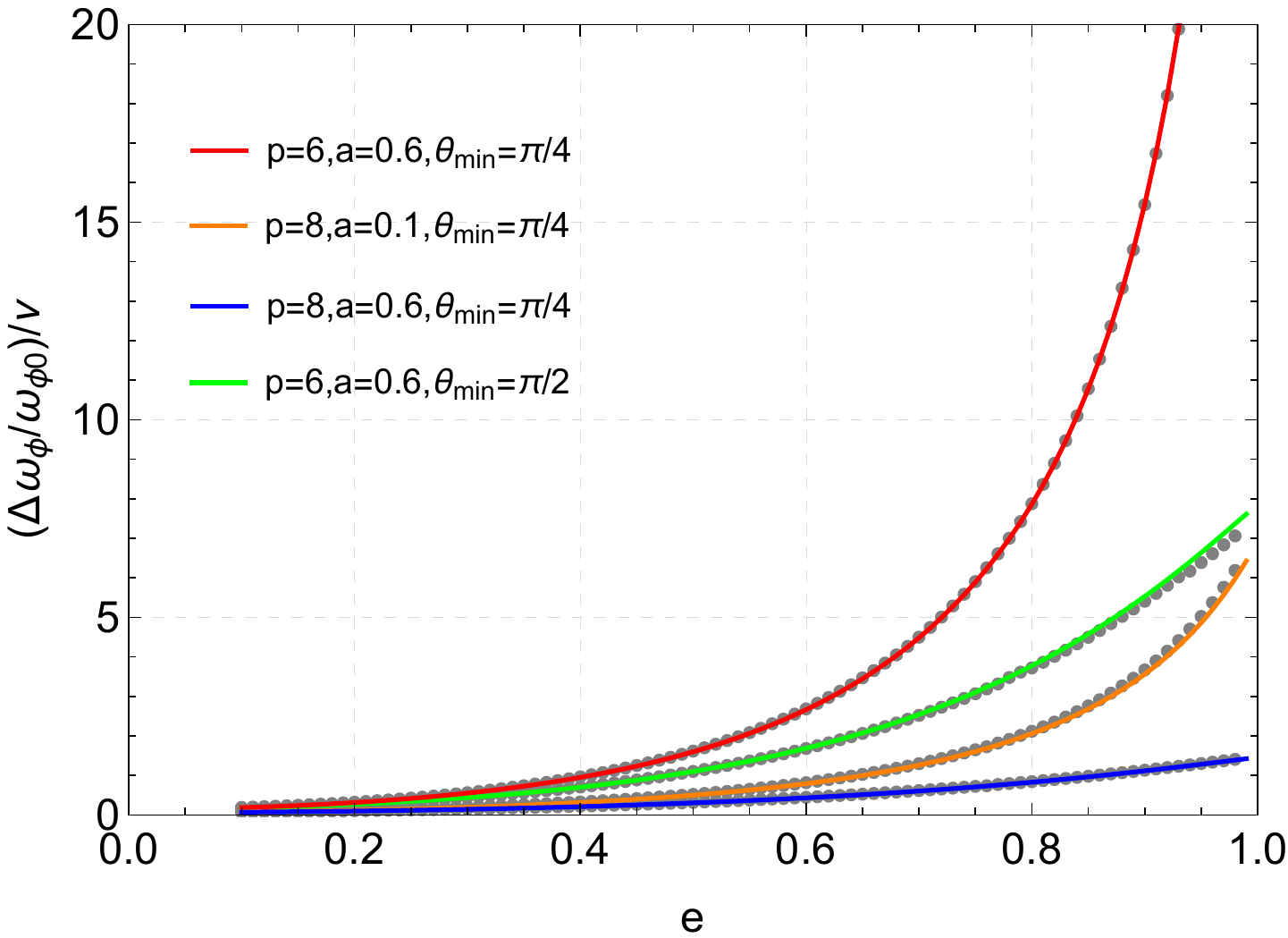}
\includegraphics[width=0.5\columnwidth]{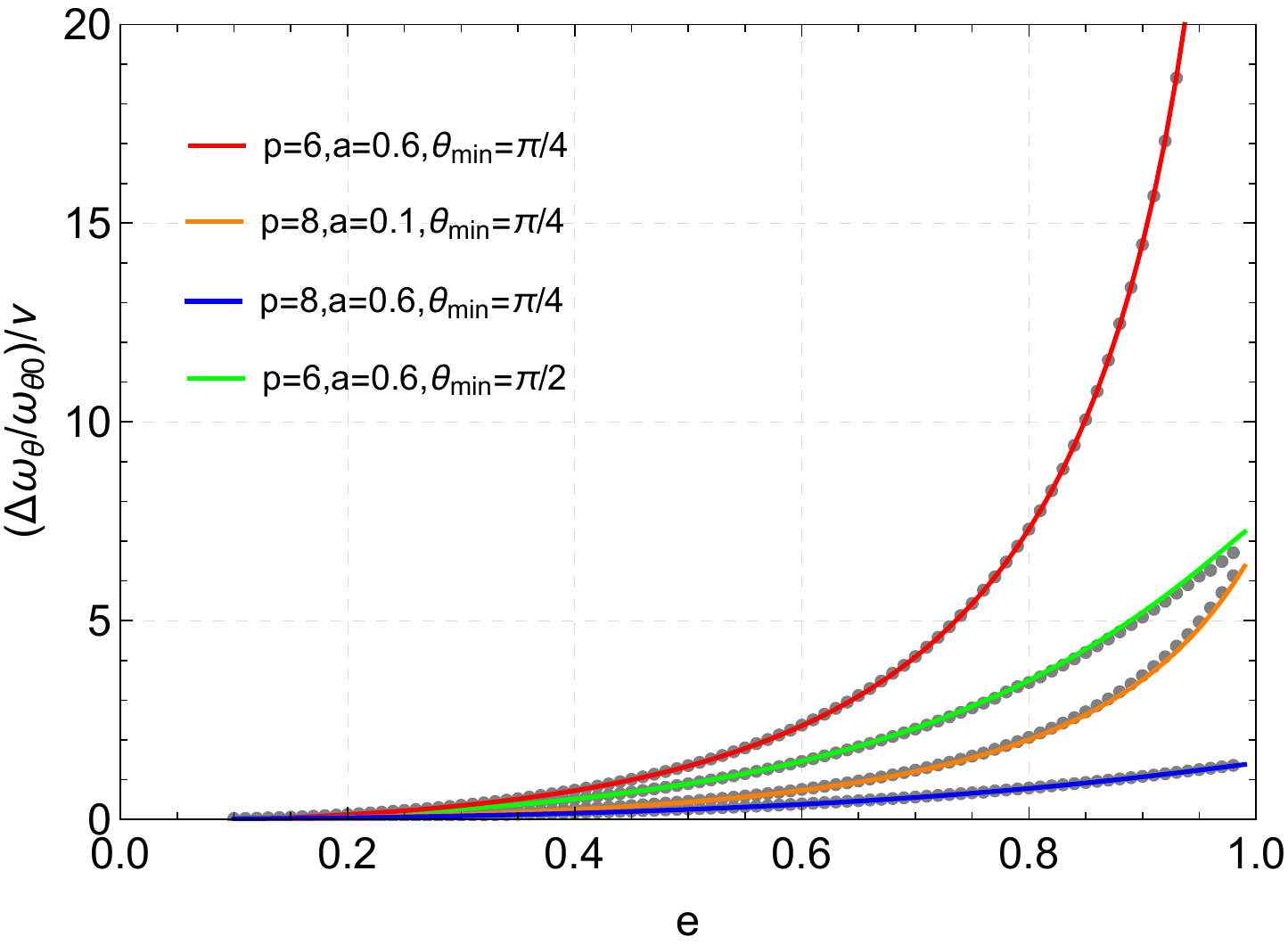}
\caption{\emph{Frequency shifts $\frac{\Delta\omega}{\nu\omega_{0}}$ vs eccentricity $e$ in the cases of various $a,~\nu,~ p,~ \theta_{\rm min}$.} The solid line and points represent $\nu =10^{-4}$ and $10^{-6}$ respectively.}
\label{fig:omega_e}
\end{center}
\end{figure}
\comment{
\begin{figure}[!h]
\begin{center}
\includegraphics[width=0.5\columnwidth]{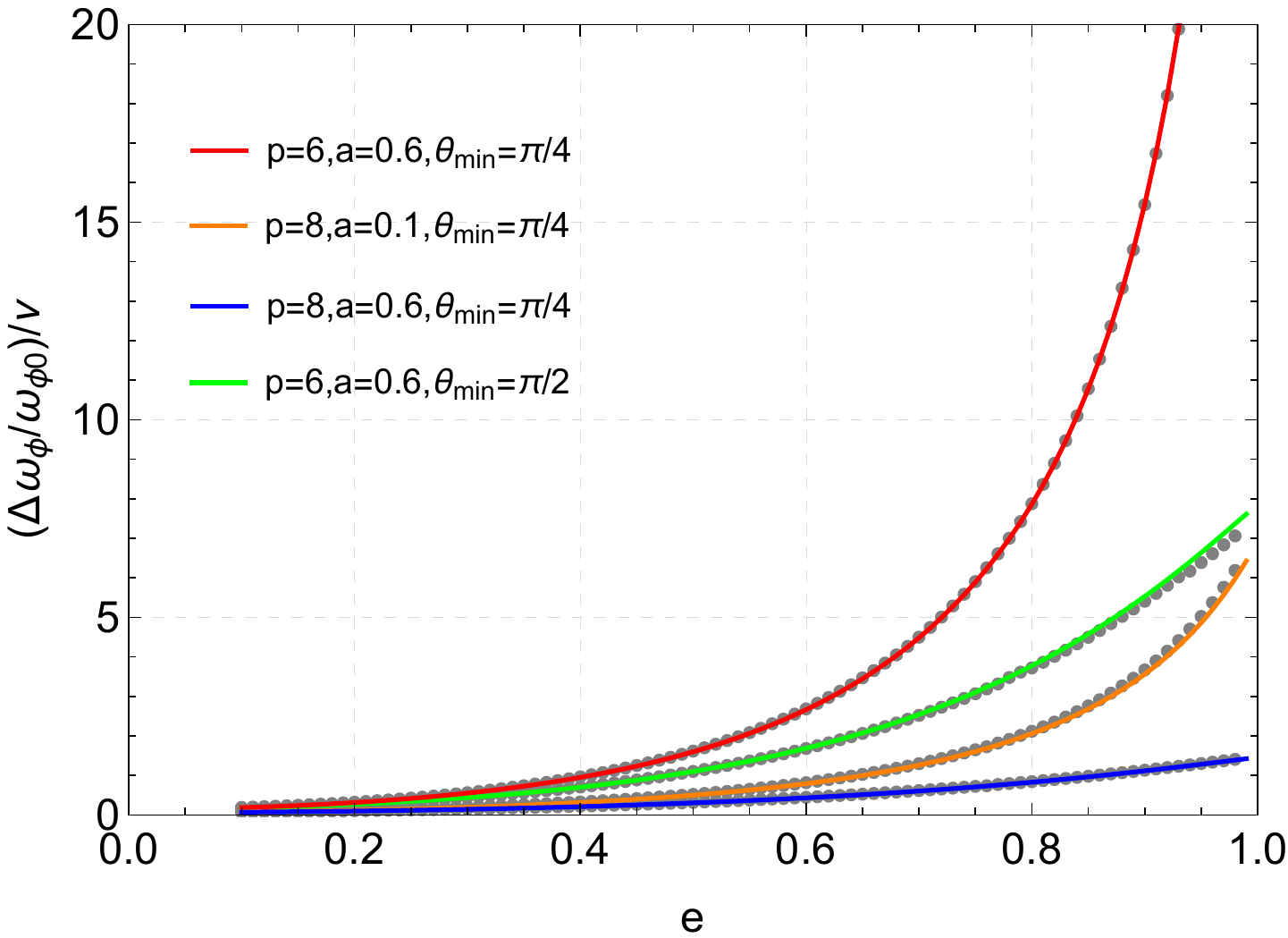}
\caption{\emph{Frequency shifts $\frac{\Delta\omega_{\phi}}{\nu\omega_{\phi0}}$ vs eccentricity $e$ in the cases of various $a,~\nu,~ p,~ \theta_{\rm min}$.} The solid line and points represent $\nu =10^{-4}$ and $10^{-6}$ respectively.}
\label{fig:6}
\end{center}
\end{figure}
\begin{figure}[!h]
\begin{center}
\includegraphics[width=0.5\columnwidth]{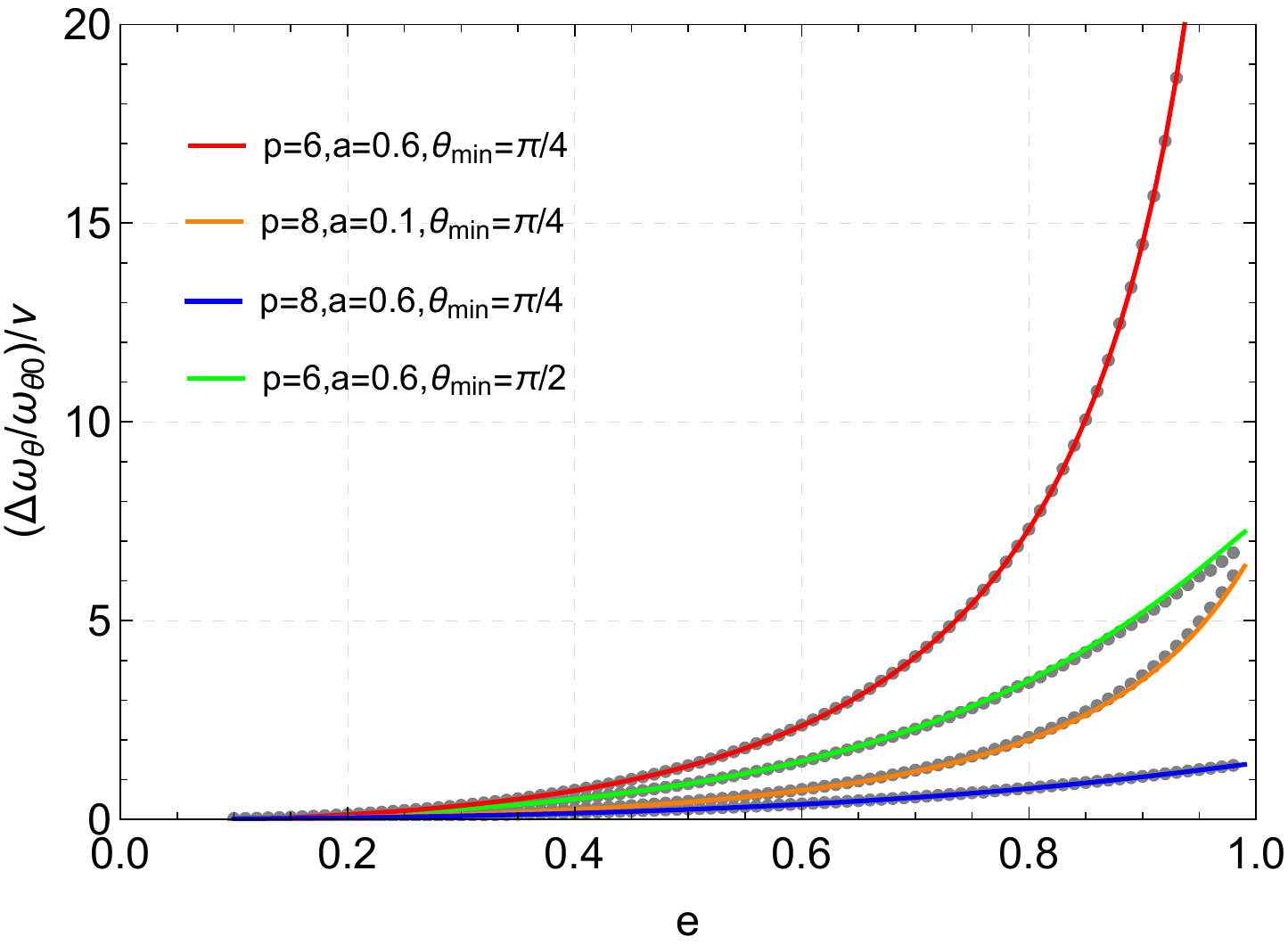}
\caption{\emph{Frequency shifts $\frac{\Delta\omega_{\theta}}{\nu\omega_{\theta0}}$ vs eccentricity $e$ in the cases of various $a,~\nu,~ p,~ \theta_{\rm min}$.} The solid line and points represent $\nu =10^{-4}$ and $10^{-6}$ respectively.}
\label{fig:7}
\end{center}
\end{figure}
}

\begin{figure}[!h]
\begin{center}
\includegraphics[width=0.5\columnwidth]{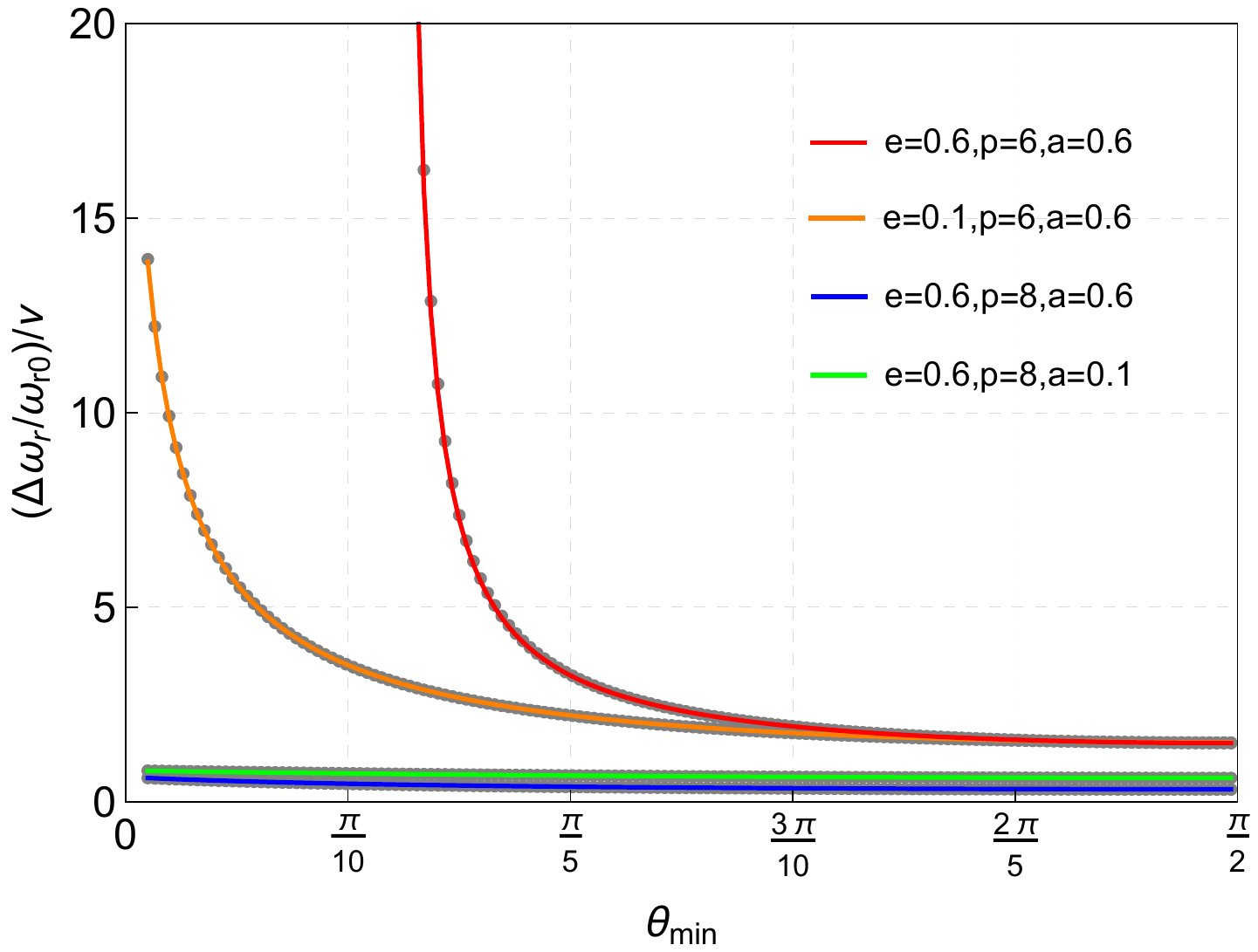}
\includegraphics[width=0.5\columnwidth]{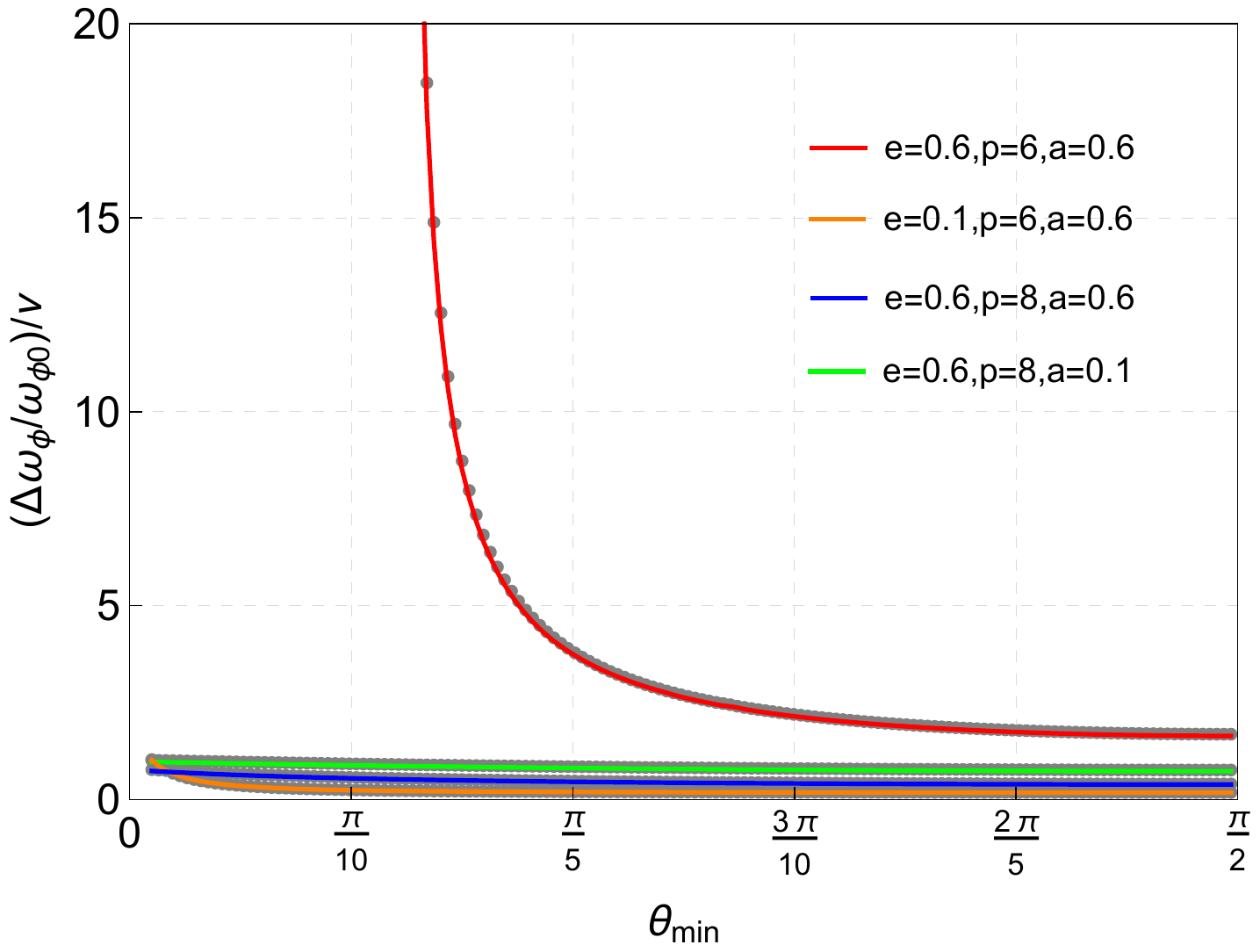}
\includegraphics[width=0.5\columnwidth]{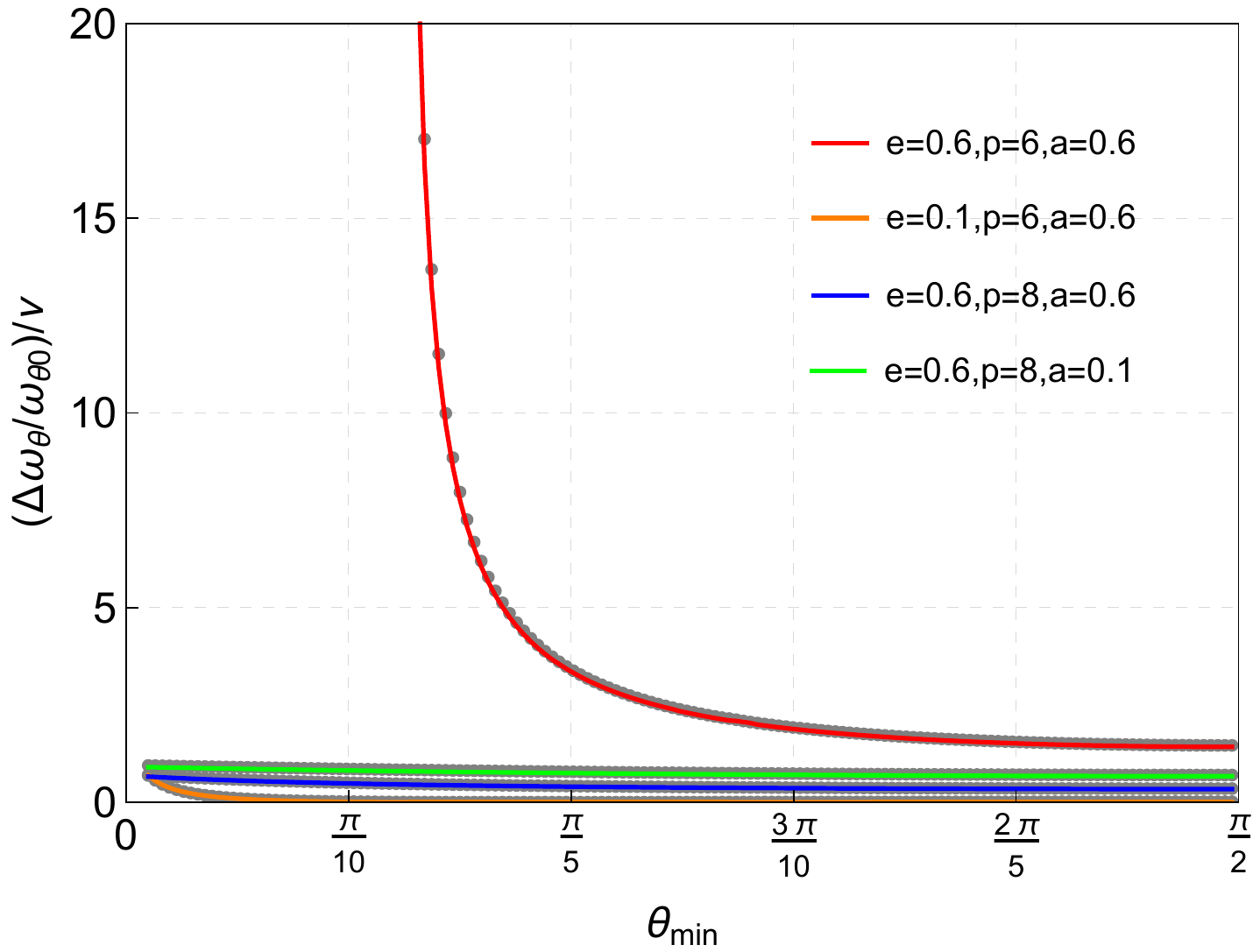}
\caption{\emph{Frequency shifts $\frac{\Delta\omega}{\nu\omega_{0}}$ vs the minimal polar angle $\theta_{\rm min}$ in the cases of various $a,~\nu,~ p,~ e$.} The solid line and points represent $\nu =10^{-4}$ and $10^{-6}$ respectively.}
\label{fig:omega_i}
\end{center}
\end{figure}
\comment{
\begin{figure}[!h]
\begin{center}
\includegraphics[width=0.5\columnwidth]{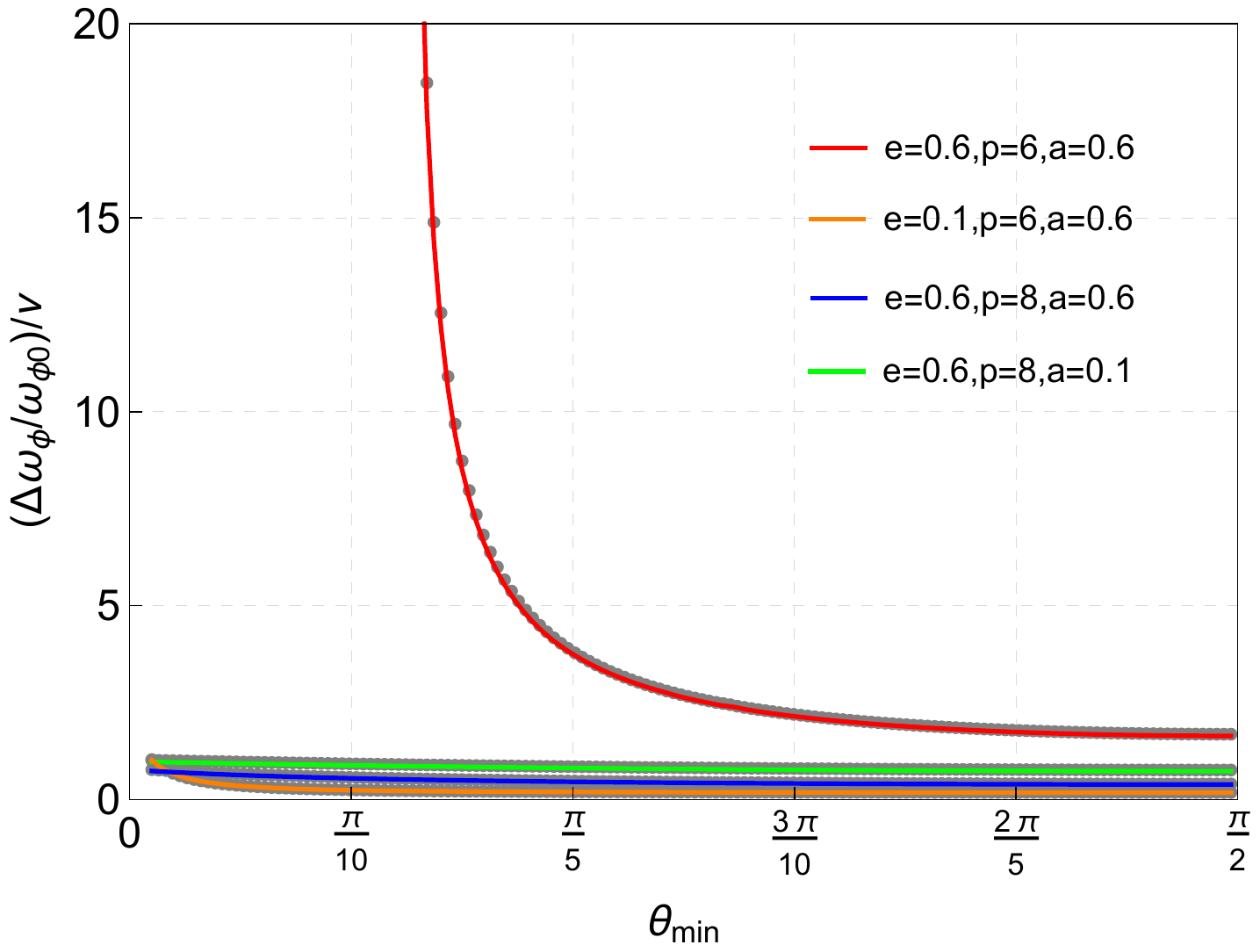}
\caption{\emph{Frequency shifts $\frac{\Delta\omega_{\phi}}{\nu\omega_{\phi0}}$ vs the minimal polar angle $\theta_{\rm min}$ in the cases of various $a,~\nu,~ p,~ e$.} The solid line and points represent $\nu =10^{-4}$ and $10^{-6}$ respectively.}
\label{fig:9}
\end{center}
\end{figure}
\begin{figure}[!h]
\begin{center}
\includegraphics[width=0.5\columnwidth]{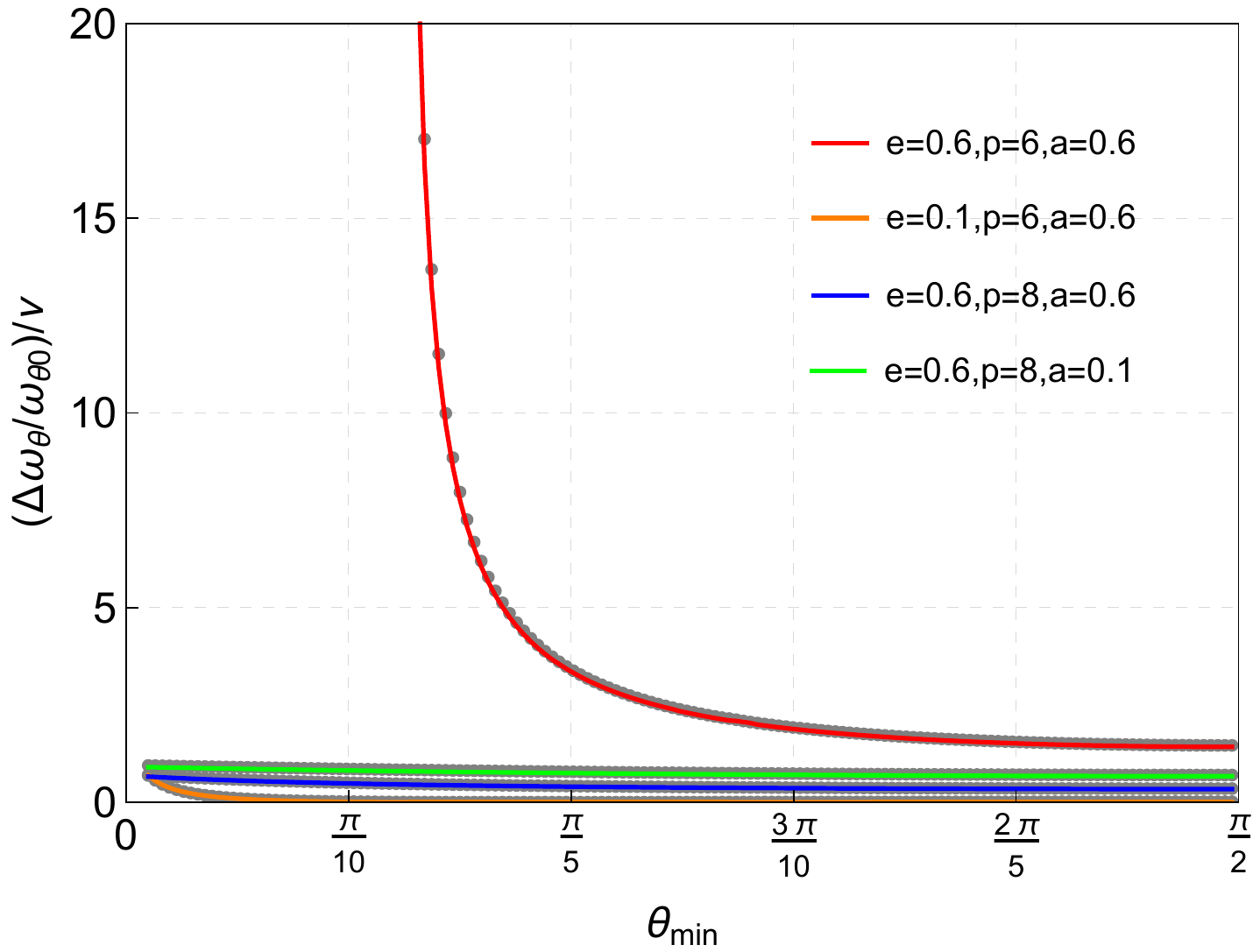}
\caption{\emph{Frequency shifts $\frac{\Delta\omega_{\theta}}{\nu\omega_{\theta0}}$ vs the minimal polar angle $\theta_{\rm min}$ in the cases of various $a,~\nu,~ p,~ e$.} The solid line and points represent $\nu =10^{-4}$ and $10^{-6}$ respectively.}
\label{fig:10}
\end{center}
\end{figure}
}

\begin{figure}[!h]
\begin{center}
\includegraphics[width=0.5\columnwidth]{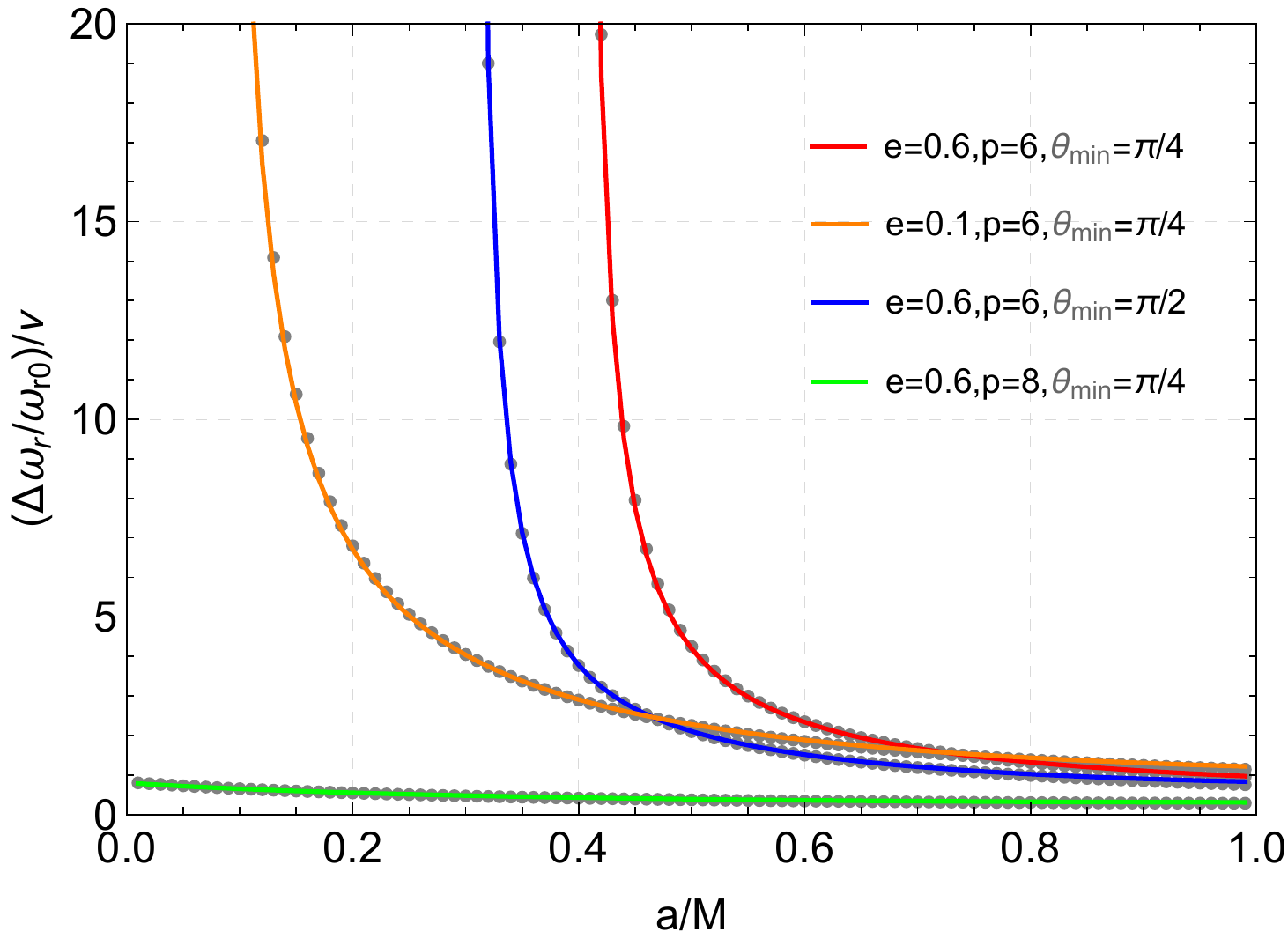}
\includegraphics[width=0.5\columnwidth]{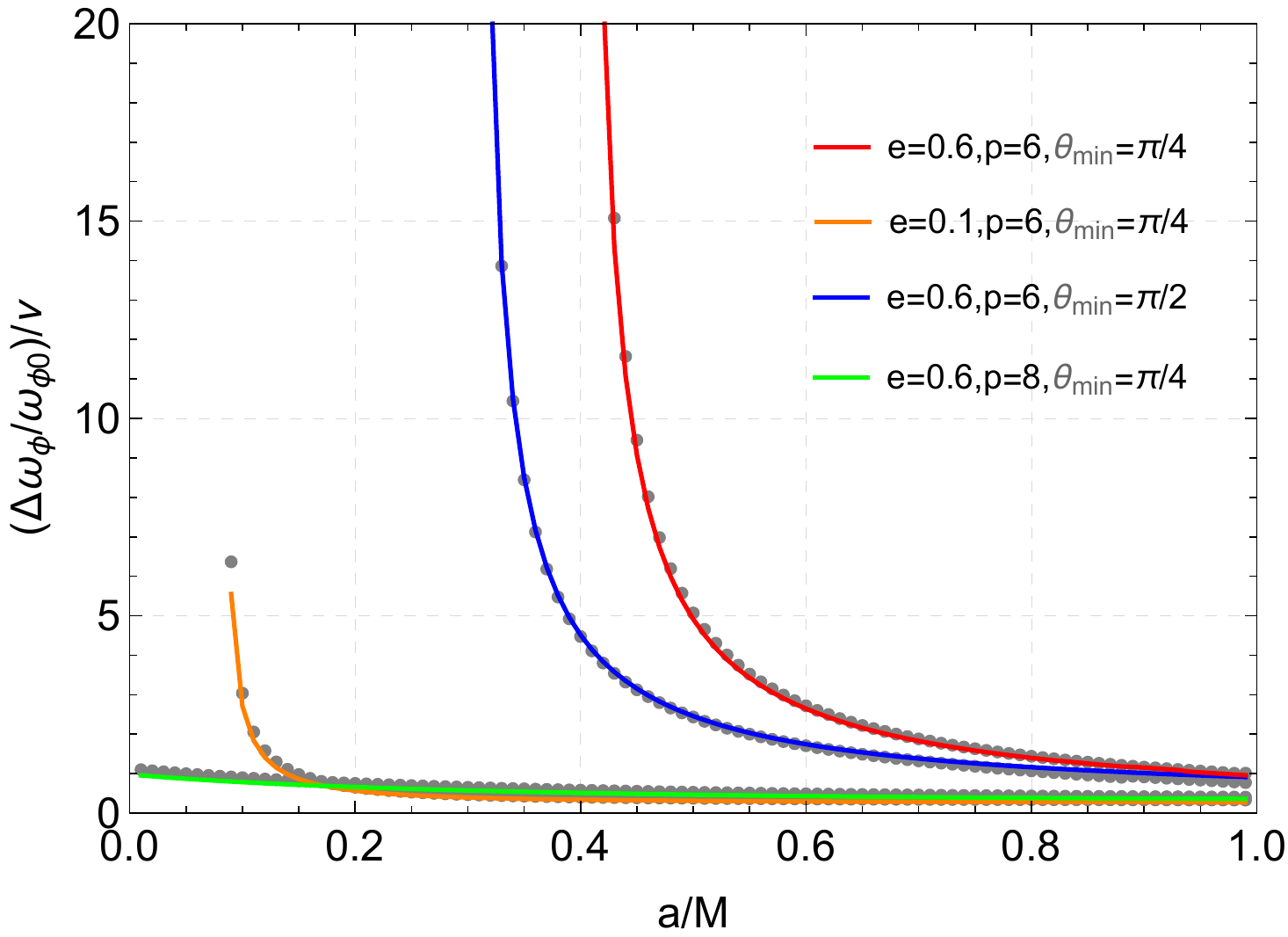}
\includegraphics[width=0.5\columnwidth]{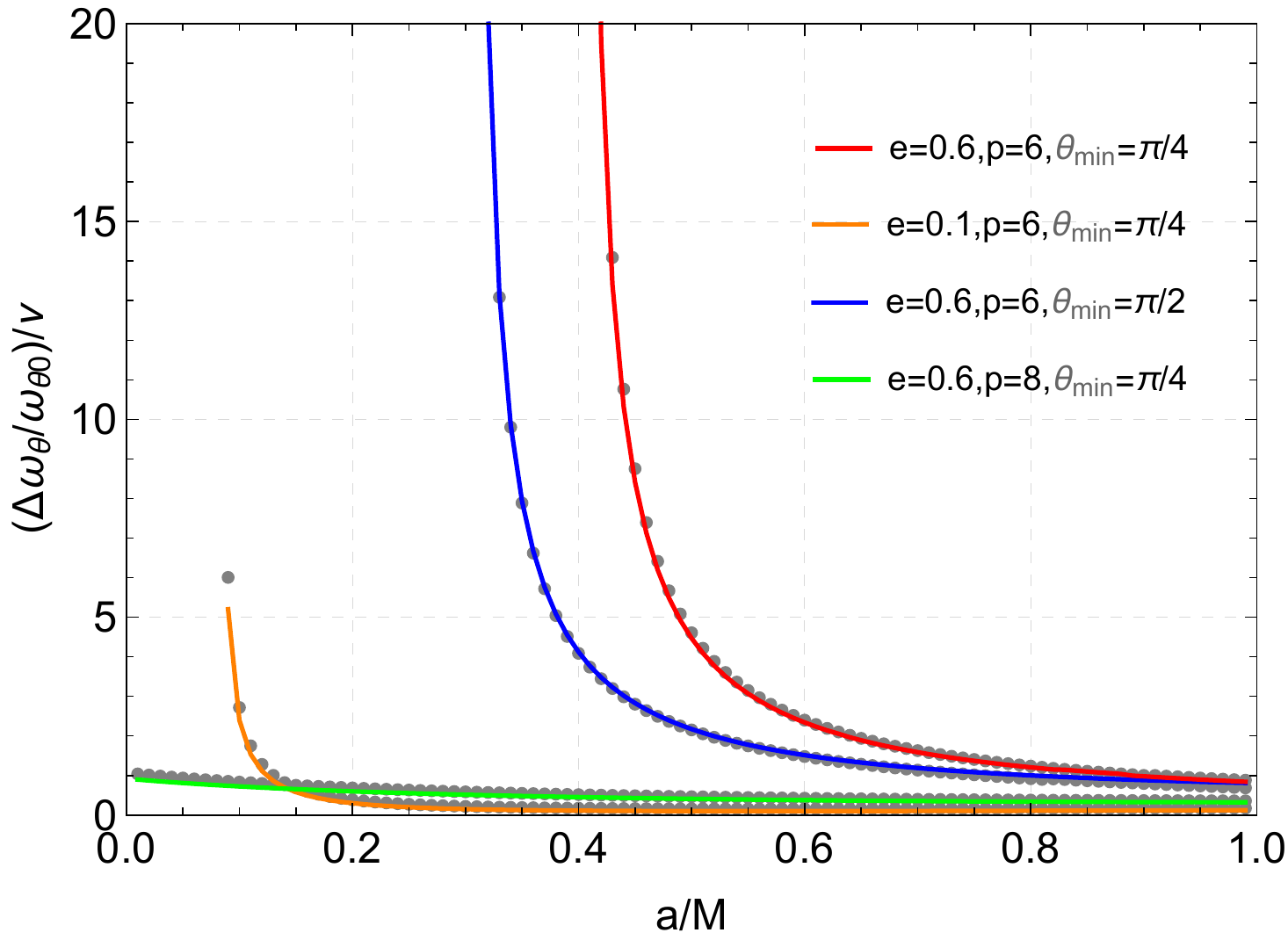}
\caption{\emph{Frequency shifts $\frac{\Delta\omega}{\nu\omega_{0}}$ vs spin $a$ in the cases of various $~\nu, ~ p, ~ e,~ \theta_{\rm min}$.} The solid line and points represent $\nu =10^{-4}$ and $10^{-6}$ respectively.}
\label{fig:omega_a}
\end{center}
\end{figure}

\comment{
\begin{figure}[!h]
\begin{center}
\includegraphics[width=0.5\columnwidth]{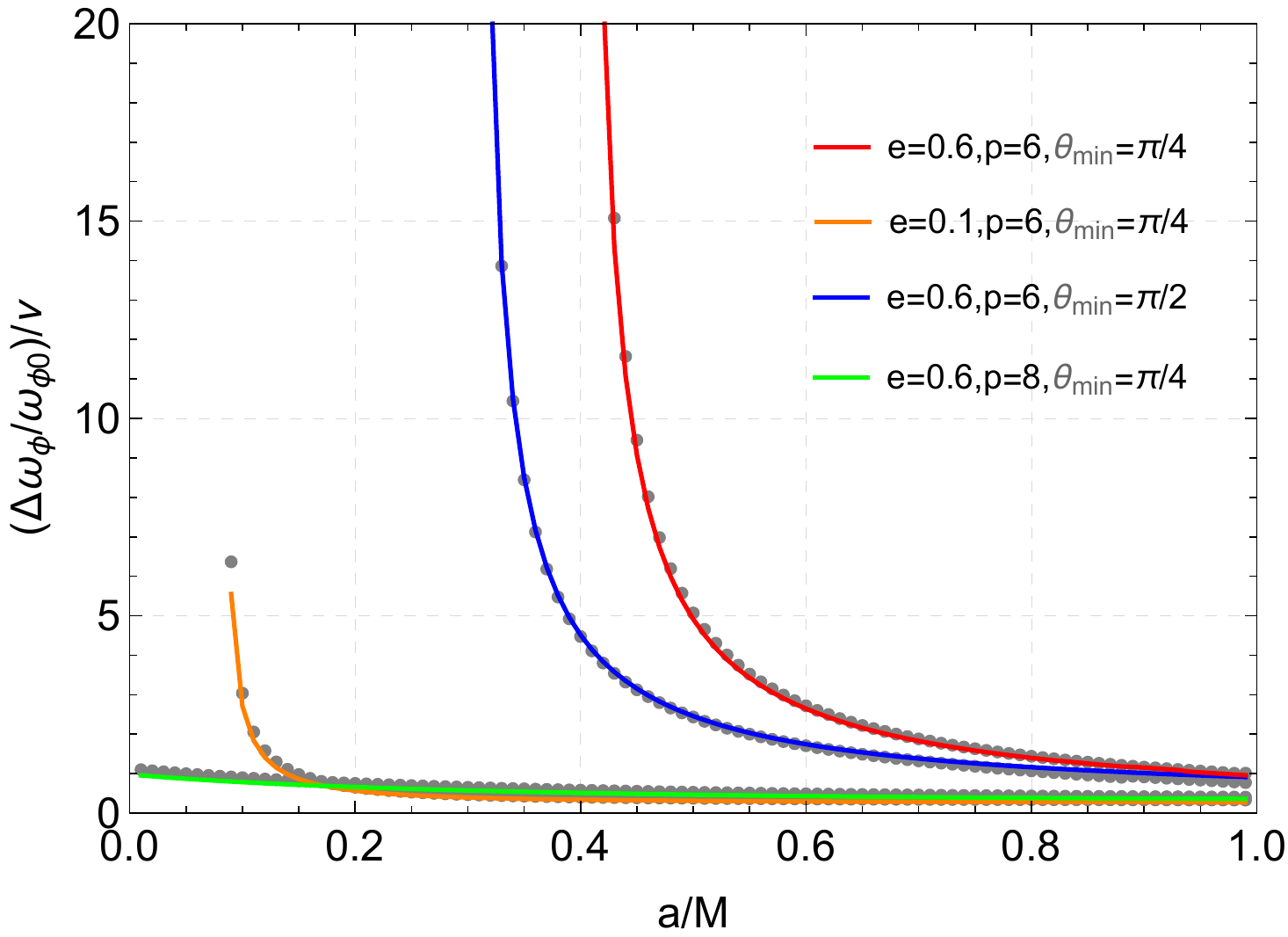}
\caption{\emph{Frequency shifts $\frac{\Delta\omega_{\phi}}{\nu\omega_{\phi0}}$ vs spin $a$ in the cases of various $~\nu, ~ p, ~ e,~ \theta_{\rm min}$.} The solid line and points represent $\nu =10^{-4}$ and $10^{-6}$ respectively.}
\label{fig:12}
\end{center}
\end{figure}
\begin{figure}[!h]
\begin{center}
\includegraphics[width=0.5\columnwidth]{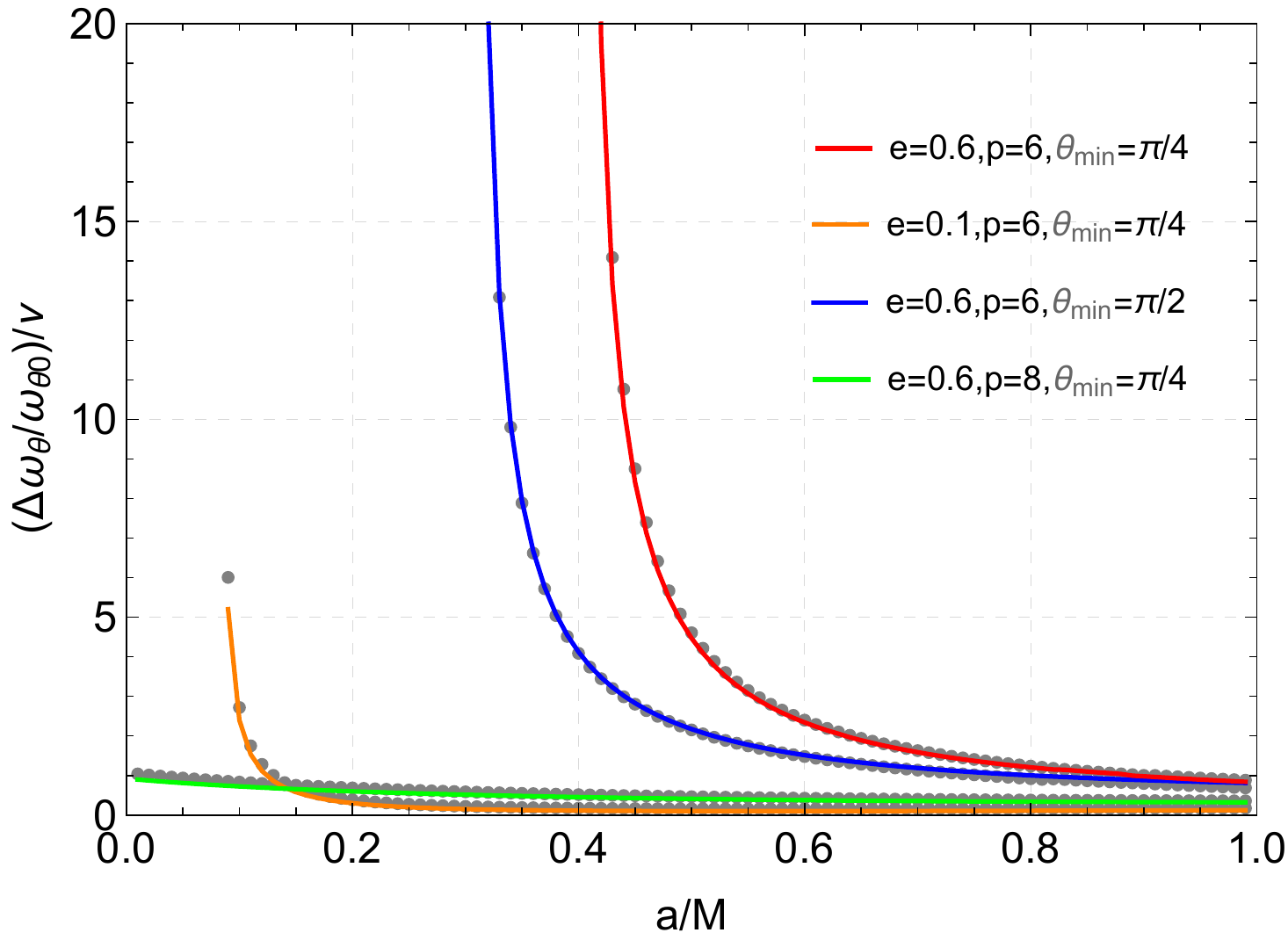}
\caption{\emph{Frequency shifts $\frac{\Delta\omega_{\theta}}{\nu\omega_{\theta0}}$ spin $a$ in the cases of various $~\nu, ~ p, ~ e,~ \theta_{\rm min}$.} The solid line and points represent $\nu =10^{-4}$ and $10^{-6}$ respectively.}
\label{fig:13}
\end{center}
\end{figure}
}

The above equations have the same forms as the test particle ones given in \cite{Schmidt2002Celestial}. However, the mass-ratio corrections have to be encoded in each variable. To explore the influence of the mass ratio on these frequencies, we
demonstrate the relative frequency shift $(\omega(\nu)-\omega_0)/\omega_0$ (where the subscript 0 means the test-particle case $\nu \rightarrow 0$) with varied orbital parameters.

In Fig.~\ref{fig:omega_p}, we can see that the three frequency shifts due to the mass ratio corrections are one order larger than the mass-ratio, especially when $p$ approaches $p_s$ (the edge of last stable orbit). The effective spin terms we omitted at
the beginning of this section have much smaller magnitudes. This proves again that the approximation we made is reasonable for EMRIs.

Fig.~\ref{fig:omega_e} demonstrates the relations of frequency shifts due to mass ratios with eccentricities. Interestingly, the azimuthal and polar frequency shifts increase when the eccentricity grows, but the radial frequency error is not sensitive to the eccentricity [about $O(\nu)$]. When $e$ becomes extreme, the shift decreases.  

The variation of frequency shifts also depends on the orbital inclination. In Fig.~\ref{fig:omega_i}, for the small semilatus rectum, the frequency errors due to mass ratio grow very fast when the orbital inclination $\iota$ increases (i.e., $\theta_{\rm min} \rightarrow 0$). However, for large $p$, the frequency shifts are not very sensitive to the orbital inclination.

Finally, Fig.~\ref{fig:omega_a} shows the frequency shifts vs the effective Kerr parameter $a$ (approachs to the Kerr parameter of the central black hole when $\nu \rightarrow 0$). When $a$ becomes smaller, the last stable orbit (LSO) is farther away from the black hole (i.e., $p_s$ becomes larger). Then for the fixed $p = 6$, the orbit is closer to the LSO as $a$ decreases, so we can see that the frequency shifts increase. When $p = 8$, which is just a little larger than the radius of the innermost stable circular orbit ($r_{\rm isco} = 6$ for a nonspinning BH), the dependence is no longer obvious 

\begin{figure}[!h]
\begin{center}
\includegraphics[width=0.6\columnwidth]{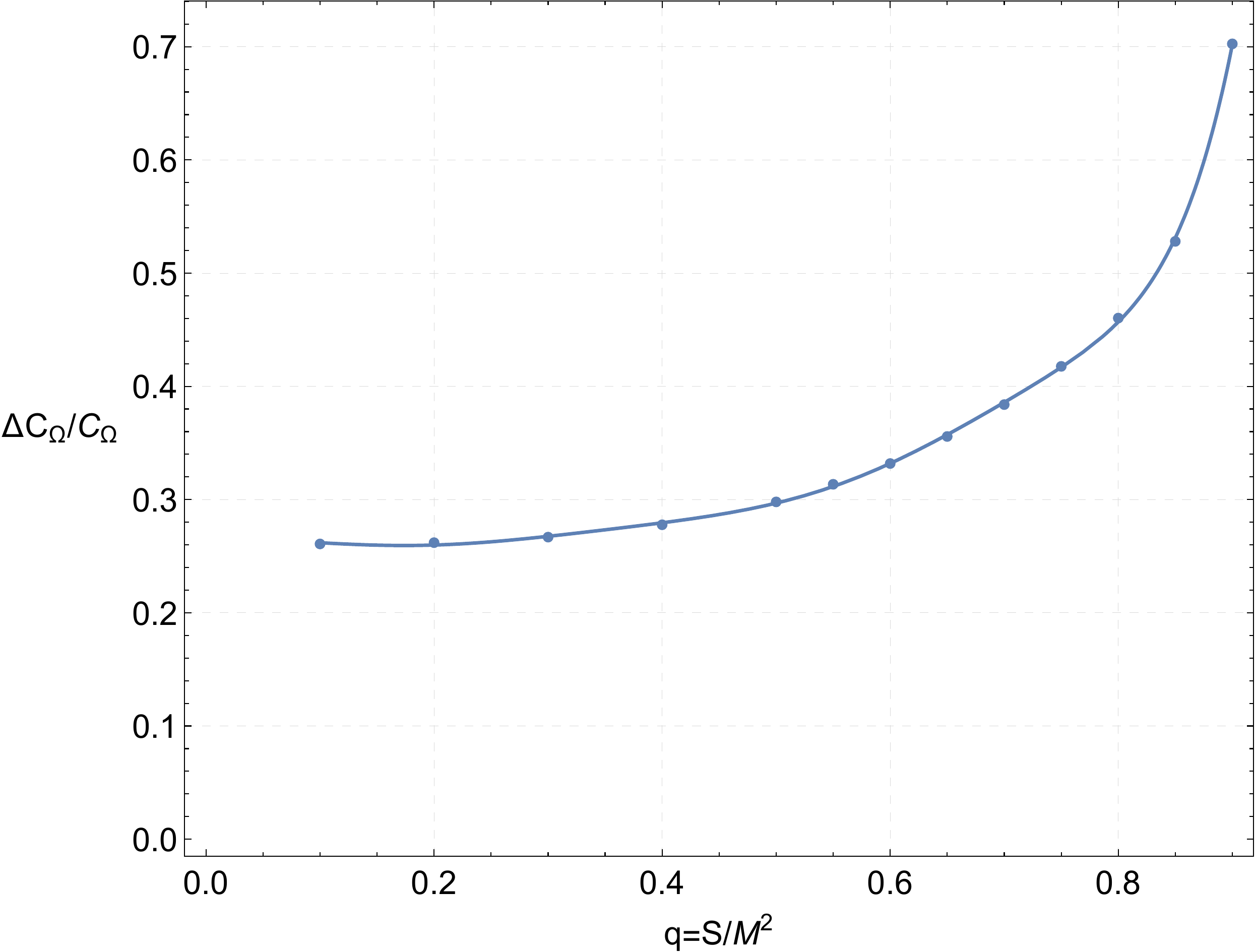}
\caption{\emph{Relative difference of  ISCO frequency shift $\Delta C_{\Omega}/C_{\Omega}:=1-C^{EOB}_{\Omega}/C_{\Omega}^{GSF}$ between the EOB and GSF methods~\rm{\cite{Isoyama2014}}.}}
\label{fig:0}
\end{center}
\end{figure}

All of the above results state that the mass-ratio correction has a substantial influence on the orbital frequencies even for the extreme-mass-ratio limit. This means that in the construction of the waveform templates for EMRIs, the mass ratio needs to be included in the orbital calculations. However, as we stated in the Introduction, the frequency shifts in Fig.~\ref{fig:omega_p}-\ref{fig:omega_a} are not guaranteed quantitatively. Comparison of innermost stable circular orbit (ISCO) shifts
between the EOB and gravitational self-force (GSF) in~\cite{Isoyama2014} shows obvious deviation, but still hints that the
frequency shift is at about the same order of mass ratio,
which coincides with our results in Fig.~\ref{fig:omega_p}-\ref{fig:omega_a}. The ISCO shifts in~\cite{Isoyama2014} were calculated based on an earlier version of the EOB potential; we recalculated the ISCO shifts based on the updated potential which is used in this work \cite{steinhoff2016Apotential}, and we found that the results improve a lot (see Fig.~\ref{fig:0}). If the spin of a SMBH is not extreme (below 0.8 based on the Fig.~\ref{fig:0}), the EOB’s ISCO frequency shifts exhibit
less than a 50
support our results in Figs.~\ref{fig:omega_p}-\ref{fig:omega_a} which qualitatively state
that the influence of the mass ratio on the conservative
dynamics of EMRIs cannot be ignored and give qualitative
magnitudes. In other words, the test-particle approximation
may not be enough for the EMRI waveform simulation, and
we expect more accurate self-force corrections.

\subsection{Orbits of conservative dynamics}
\label{sec:orbits}

We find evolution equations for $r$, $\phi$, and $\chi$ with the forms
\bes\label{eomsEOBrphiEL}\begin{align}
\dot r&=\frac{\partial E}{\partial P_{r}} \!=\!-\frac{g^{rr} \hat{P_r}}{E/M \left(g^{tt}\hat{H}_{\rm eff}-g^{t \phi}\hat{L}_z\right)}\,,  \label{eq:rdot}\\
\dot \phi&=\frac{\partial E}{\partial P_{\phi}} \!=\!\frac{g^{t \phi}-\Big[g^{tt}g^{\phi\phi}-(g^{t \phi})^2 \Big]\frac{\hat{L}_z}{g^{tt} \hat{H}_{\rm eff}-g^{t \phi}\hat{L}_z}}{g^{tt} E/M}\,,\label{eq:phidot}\\
\dot{\theta}\!&=\frac{\partial E}{\partial P_{\theta}} \!=\!-\frac{g^{\theta \theta} \hat{P_{\theta}}}{E/M \left(g^{tt}\hat{H}_{\rm eff}-g^{t \phi}\hat{L}_z\right)}\,\label{eq:thetadot}.
\end{align}
\ees
where $\hat{P}_r$, $\hat{P}_{\theta}$ have been analytically obtained in Eqs.~\eqref{eq:ptheta}-\eqref{eq:pr}, and $\hat{H}_{\rm eff}$, $\hat{L}_{z}$ have been given in Eq. \eqref{eq:psofep}. The metric components are Eqs.~(\ref{def_metric_in})-(\ref{def_metric_fin}). Due to the definitions of $\xi,~\chi$, all the variables by $r$ and $\theta$ can be transfer to the functions of $\xi,~\chi$. Finally, the above equations can now be expressed in terms of only the variables $(\xi,\chi)$ and orbital parameters $(p,e,\theta_{\rm min})$ or $(p,e,\iota)$, 
\bes\bea
\dot \xi&&=-\frac{(1+e\cos\xi)^2}{epM\sin\xi}\frac{g^{rr} \hat{P_r}}{E/M \left(g^{tt}\hat{H}_{\rm eff}-g^{t \phi}\hat{L}_z\right)} = \Xi(p,e,\iota,\xi,\chi) \,,  \label{eq:xidot}\\
\dot{\chi}\!&&=-\frac{g^{\theta \theta} \sqrt{\left(a^2 \left(1-\hat{H}_{\rm eff}^2\right)\right) \left(z_+^2-z_-^2 \cos ^2\chi \right)}}{E/M \left(g^{tt}\hat{H}_{\rm eff}-g^{t \phi}\hat{L}_z\right)} = \Theta(p,e,\iota,\xi,\chi)  \,\label{eq:chidot}, \\
\dot \phi&&=\frac{g^{t \phi}-\Big[g^{tt}g^{\phi\phi}-(g^{t \phi})^2 \Big]\frac{\hat{L}_z}{g^{tt} \hat{H}_{\rm eff}-g^{t \phi}\hat{L}_z}}{g^{tt} E/M} = \Phi(p,e,\iota,\xi,\chi) \,.\label{eq:phidot2}
\eea\ees
The detailed expressions of $\Xi,\Theta$ and $\Phi$ can be directly obtained, but they are too long to be written here. Solving
the above ordinary differential equations by numerical
integration, we can get $\xi$, $\chi$, and $\phi$ associated with coordinate time $t$, i.e., the orbital motion. Projecting the Boyer-Lindquist coordinates onto a spherical coordinate grid, we can define the corresponding Cartesian coordinate system,
\bes\label{sphere}
\bea
\tilde{x}&=&\cfrac{p \cos\phi\sqrt{1-z_-^2\cos ^2\chi }}{1+e \cos \xi}\,,\\
\tilde{y}&=&\cfrac{p \sin\phi\sqrt{1-z_-^2\cos ^2\chi }}{1+e \cos \xi}\,,\\
\tilde{z}&=&\cfrac{p z_-\cos\chi}{1+e \cos \xi}\,.
\eea
\ees

\begin{figure}[!h]
\centering
\subfloat[\;$a=0.5~M,\;e=0.2,\;p=5~M,\;p_s=4.99~M$]{      
\begin{minipage}[c]{.5\linewidth}
\centering
\includegraphics[width=\textwidth]{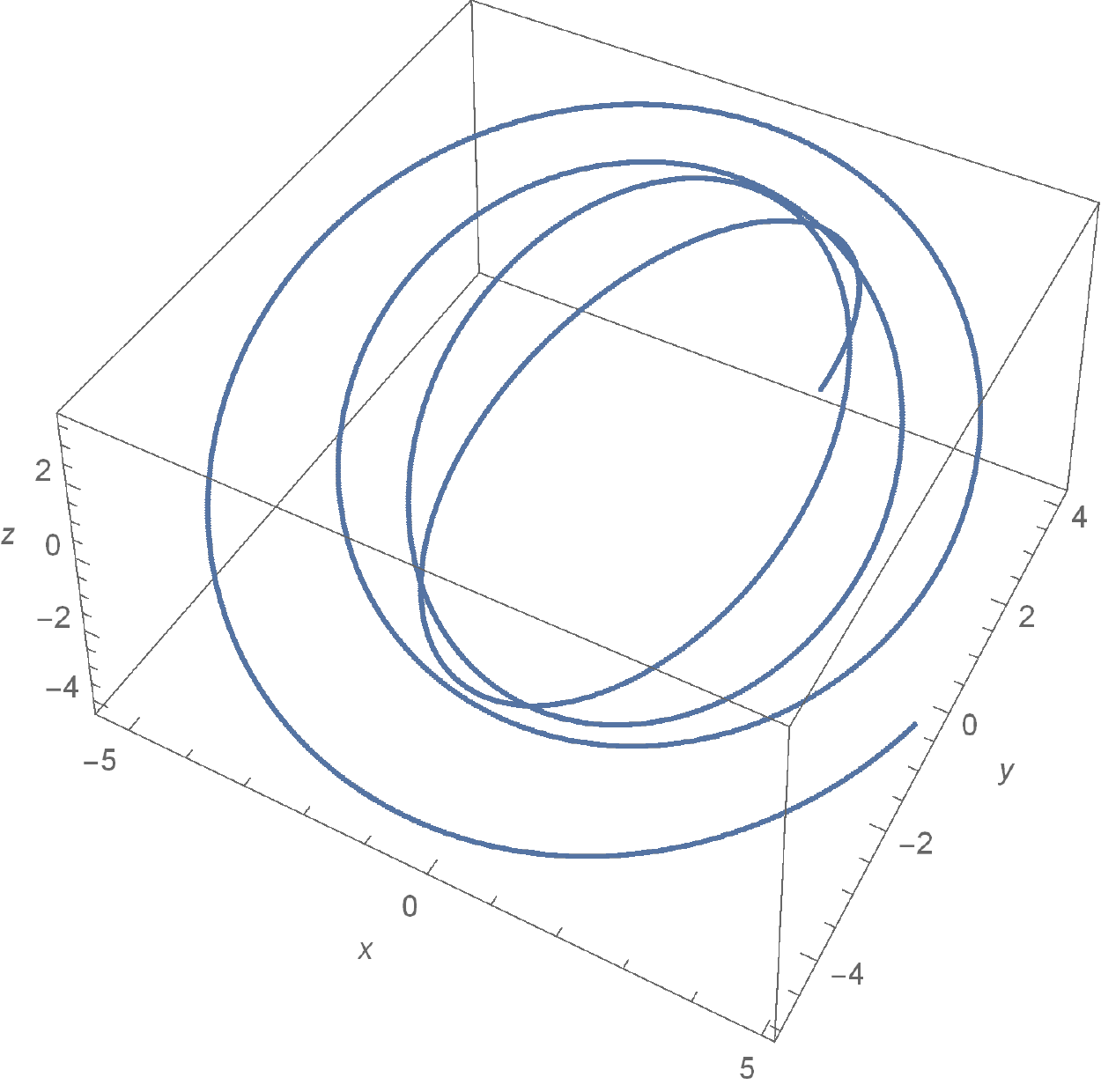}
\end{minipage}
}
\subfloat[\;$a=0.5~M,\;e=0.1,\;p=5~M,\;p_s=4.84~M$]{
\begin{minipage}[c]{.5\linewidth}
\centering
\includegraphics[width=\textwidth]{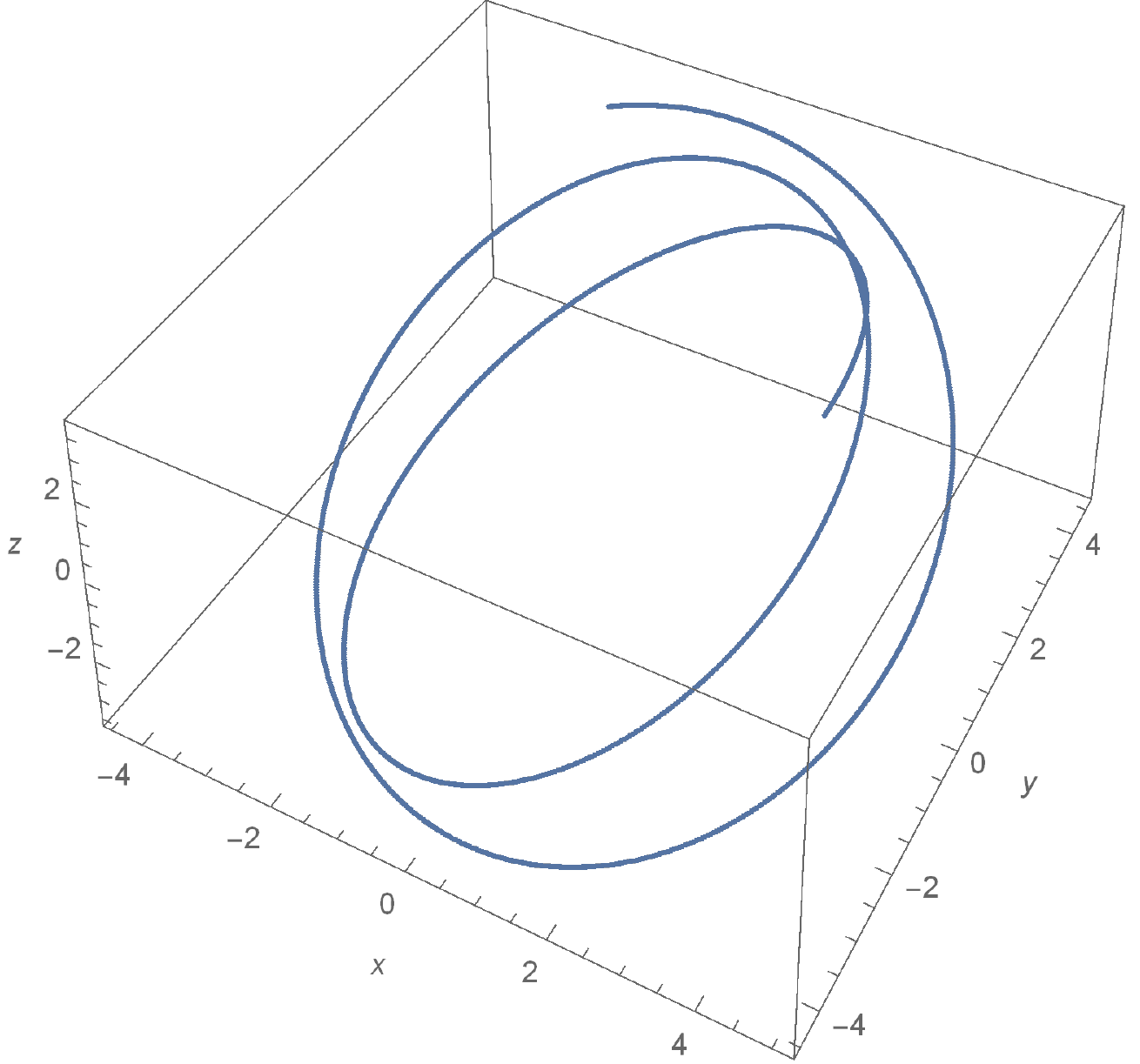}
\end{minipage}
}\\
\subfloat[\;$a=0.6~M,\;e=0.6,\;p=6~M,\;p_s=5.28~M$]{      
\begin{minipage}[c]{.5\linewidth}
\centering
\includegraphics[width=\textwidth]{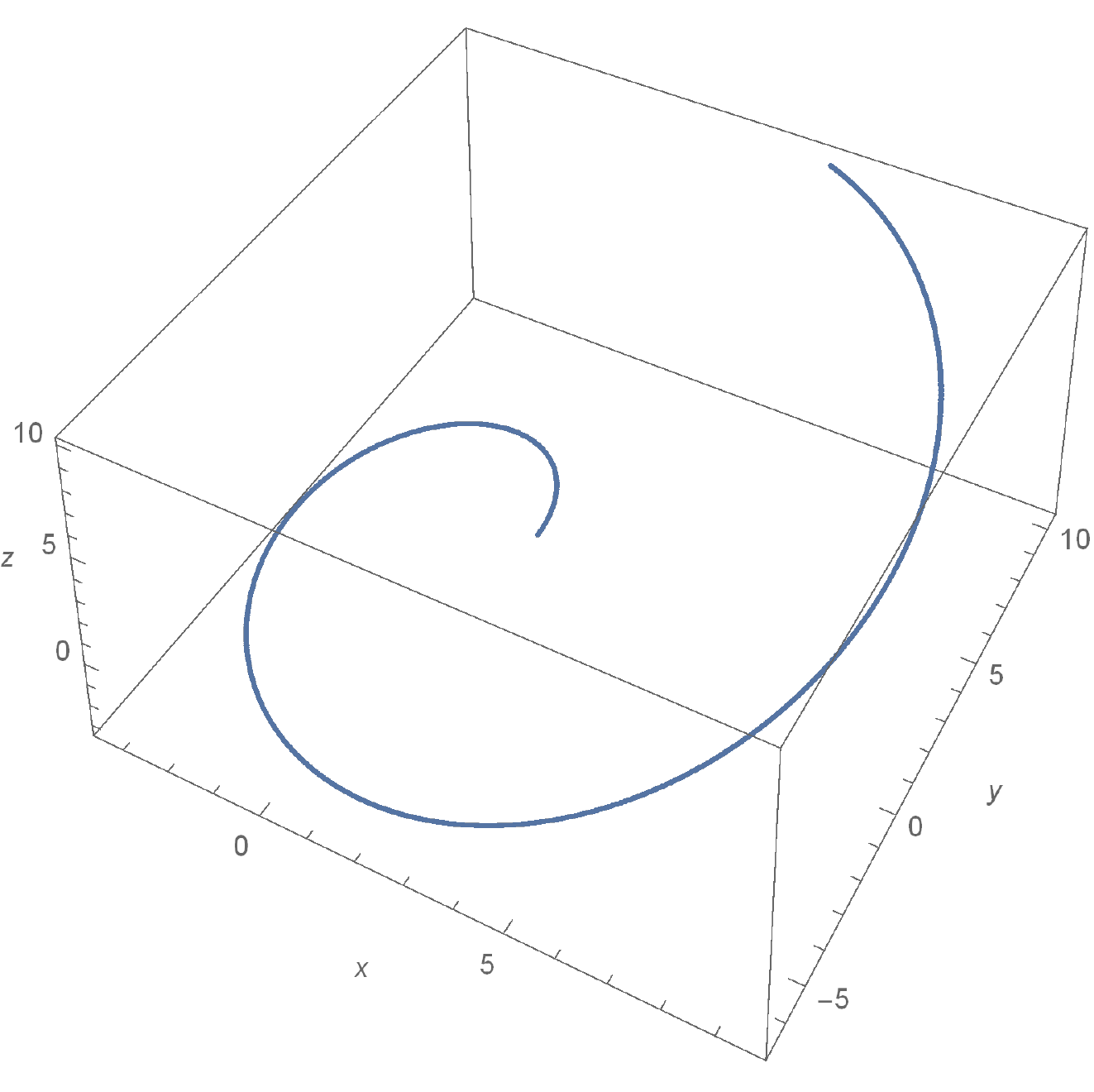}
\end{minipage}
}
\subfloat[\;$a=0.6~M,\;e=0.1,\;p=6~M,\;p_s=4.51~M$]{
\begin{minipage}[c]{.5\linewidth}
\centering
\includegraphics[width=\textwidth]{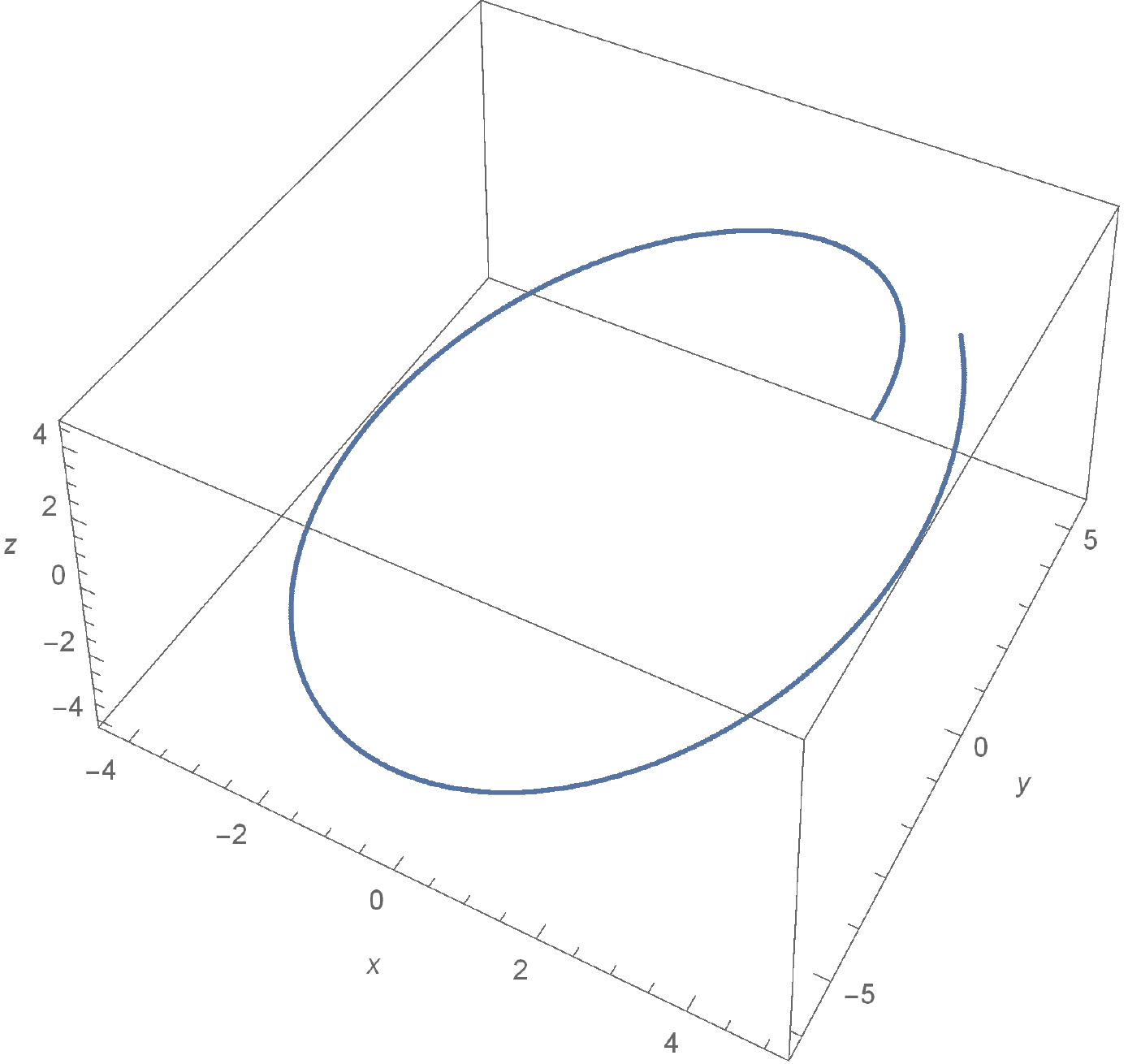}
\end{minipage}
}
\caption{\emph{Four orbits of different initial parameters}. We set $\nu=10^{-4},\theta_{\rm min}=\pi/4$ in the all four orbits and just plot half orbital period from  the apastron to the periastron.}
\label{fig:14}
\end{figure}

Combining the equations of the motion (\ref{eq:xidot}) - (\ref{eq:phidot2}) we can plot the orbits of conservative dynamics in the Cartesian coordinates (see Fig.~\ref{fig:14}). 

Like the zoom-whirl orbits \cite{glampedakis2002zoom} in $r$-$\phi$ components, from Fig.~\ref{fig:14} there are similar zoom-whirl phenomena in the $r$-$\theta$ directions. Different orbital parameters with a corresponding prograde separatrix have different polar angular periods. We can imagine that as the orbit gradually approaches the separatrix, the test particle will spend a more considerable amount of its orbital ‘‘life’’ (in both the azimuthal and polar directions) close to the periastron. 

\comment{
\subsection{The shift of last stable orbits due to mass ratio}
\label{sec:lso}
In test particle limit $\hat{P}_r=0$ has four distinct real roots, expanding $\Delta^2\hat{P}_r^2$ as 
\be
\Delta^2\hat{P}_r^2=(\hat{H}_{\rm eff}^2-1)(r-r_1)(r-r_2)(r-r_3)(r-r_4)
\ee
where $\Delta=r^2-2Mr+a^2$, then the third root $r_3$ and the fourth root $r_4$ are found to be
\bea
r_3=\frac{p^2 \left(1-e^2-p \left(1-\hat{H}_{\rm eff}^2\right)\right)+\sqrt{p^4 \left(1-e^2-p \left(1-\hat{H}_{\rm eff}\right)\right)^2+a^2 p^2 \hat{Q} \left(1-e^2\right)^3\left(1-\hat{H}_{\rm eff}^2\right)}}{p^2 \left(1-e^2\right) \left(1-\hat{H}_{\rm eff}^2\right)}\,\\
r_4=\frac{p^2 \left(1-e^2-p \left(1-\hat{H}_{\rm eff}^2\right)\right)-\sqrt{p^4 \left(1-e^2-p \left(1-\hat{H}_{\rm eff}\right)\right)^2+a^2 p^2 \hat{Q} \left(1-e^2\right)^3\left(1-\hat{H}_{\rm eff}^2\right)}}{p^2 \left(1-e^2\right) \left(1-\hat{H}_{\rm eff}^2\right)}
\eea
stable bound orbits should satisfy $r_3<r_2$
}
\subsection{Evolution of orbital parameters under radiation reaction}
\label{sec:evolution}
The description of geodesic motion around BHs is based on the semilatus rectum $p$, the eccentricity $e$ and the orbital inclination $\iota$, together with three phase variables associated with the spatial geometry of the radial, azimuthal and polar motion denoted by $(\xi,\phi,\chi)$. $\xi$ was already defined by expressing the radial motion as 
\be
r=\frac{p M}{1+e\cos\xi}\,.\\ \label{eq:rofxi}
\ee
Taking the derivation of Eq.~(\ref{eq:rofxi}), we get the evolution equation for the phase variable $\xi$,
\be
\label{eq:xidotgen1}
\dot \xi=\frac{(1+e\cos\xi)^2}{epM\sin\xi}\dot r-\frac{1+e\cos\xi}{ep\sin\xi}\dot p+\frac{\cot\xi}{e}\dot e.
\ee
For a conservative system, $\dot{p}=\dot{e}=0$. However, if we take the radiation reaction of GWs into account, the rate of change of the energy,  the reduced angular momentum, and the Carter constant are given by
\bes
\label{eq:energydot}
\begin{align}
\frac{dE}{dt}&=\frac{\partial E}{\partial p}\dot p+\frac{\partial E}{\partial e}\dot e+\frac{\partial E}{\partial  \iota}\dot \iota\,,\label{eq:energydot1}\\
\frac{d \hat{L}_z}{dt}&=\frac{\partial \hat{L}_z}{\partial p}\dot p+\frac{\partial \hat{L}_z}{\partial e}\dot e+\frac{\partial \hat{L}_z}{\partial \iota}\dot \iota\,,\\
\frac{d\hat{Q}}{dt}&=\frac{\partial \hat{Q}}{\partial p}\dot p+\frac{\partial \hat{Q}}{\partial e}\dot e+\frac{\partial \hat{Q}}{\partial \iota}\dot \iota\,.\label{eq:energydot3}
\end{align}
\ees
From Eqs.~(\ref{eq:energydot1})-(\ref{eq:energydot3}), we express the evolution of $(p,~e,~\iota)$ using the energy and angular momentum fluxes of gravitational radiation,
\begin{widetext}
\bes
\bea
\dot{p}&=c_{(\hat{L}_z,\hat{Q})(e,\iota)}\frac{dE}{dt}+c_{(E,\hat{Q})(\iota,e)}\frac{d\hat{L}_z}{dt}+c_{(E,\hat{L}_z)(e,\iota)}\frac{d\hat{Q}}{dt}\,\label{eq:edot}\\
\dot{e}&=c_{(\hat{L}_z,\hat{Q})(\iota,p)}\frac{dE}{dt}+c_{(E,\hat{Q})(p,\iota)}\frac{d\hat{L}_z}{dt}+c_{(E,\hat{L}_z)(\iota,p)}\frac{d\hat{Q}}{dt}\,\label{eq:pdot}\\
\dot{\iota}&=c_{(\hat{L}_z,\hat{Q})(p,e)}\frac{dE}{dt}+c_{(E,\hat{Q})(e,p)}\frac{d\hat{L}_z}{dt}+c_{(E,\hat{L}_z)(p,e)}\frac{d\hat{Q}}{dt},\label{eq:iotadot}
\eea
where the coefficients are given by
\be
\label{eq:cAbdef}
c_{(C_1,C_2)(o_1,o_2)}=\frac{\Big[\frac{\partial C_1}{\partial o_1}\frac{\partial C_2}{\partial o_2}-\frac{\partial C_1}{\partial o_2}\frac{\partial C_2}{\partial o_1}\Big]}{\Big[\frac{\partial E}{\partial \iota}\frac{\hat{L}_z}{\partial p}-\frac{\partial E}{\partial p}\frac{\partial \hat{L}_z}{\partial \iota}\Big]\frac{\partial \hat{Q}}{\partial e}+\Big[\frac{\partial E}{\partial e}\frac{\hat{L}_z}{\partial\iota}-\frac{\partial E}{\partial \iota}\frac{\partial \hat{L}_z}{\partial e}\Big]\frac{\partial \hat{Q}_\phi}{\partial p}+\Big[\frac{\partial E}{\partial p}\frac{\hat{L}_z}{\partial e}-\frac{\partial E}{\partial e}\frac{\partial \hat{L}_z}{\partial p}\Big]\frac{\partial \hat{Q}_\phi}{\partial \iota}} \,,\\
\ee
\ees
\end{widetext}
where $C=\{E,\hat{L}_z,\hat{Q}\}$ and $o=\{p,e,\iota\}$, and the derivatives can be computed from the expressions in Eqs.~(\ref{eq:carter}) and (\ref{eq:psofep}). We do not write the complete expressions here because they can be calculated quite directly but are very long. 

Once we have the GW fluxes $\dot{E},~\dot{L}_z$, and $\dot{Q}$, the orbital evolution can be obtained. We have no plan to introduce the detailed fluxes in the present paper, and thus we do not calculate the specific orbital evolution here. We will leave this task to the next paper on gravitational waveforms. In this work, we just focus on the geometrization of the EOB formalism in extreme-mass-ratio cases.

The final set of EOB equations of motion with radiation reaction are Eqs.\eqref{eq:edot}-\eqref{eq:pdot} together with the evolution of the phases described by Eqs.~(\ref{eq:xidotgen1}) and (\ref{eq:iotadot}), and the radius of motion at any arbitrary moment is given by Eq.~(\ref{eq:rdot}). Now all the equations of motion are expressed in terms of only the geometric parameters $(p,e,\iota,\xi,\chi)$ and the effective Kerr parameter $a$.  

\section{Conclusions and Outlook}
\label{sec:conclusion}

In the present paper, based on the EOB deformed metric
and Hamiltonian, we gave the geometrized formalism of
the equations of motion for the inclined-eccentric EMRIs
with spinning black holes. The solutions were derived with the geometric parameters  $p$, $e$ and $\iota$ instead of the EOB coordinates and momentum. The fundamental properties of the motion due to the mass ratio and black hole spin were discussed. We also gave expressions for three orbital frequencies $\omega_r, ~\omega_\theta, ~\omega_\phi$. With these formalisms in hand, it is convenient to obtain the motion of a compact
object around a supermassive black hole with the orbital
parameters  $p,e,\iota$.  

Our results show that the influence on the orbital motion due to the small compact object's gravitational self-force on the background of a SMBH cannot be ignored. The analytical formalism in this work makes the inclusion of the mass ratio in the motion much more intuitive. Though we do not give a waveform template in the present work, We believe that our analytical method (not the formalism themselves) should be an useful way to build waveform models in the future for EMRIs to replace the test-particle approximation which is used in popular waveform models.  

In the present work, a few approximations have been used. As we mentioned before, in the present model we temporarily omitted the effective spin of the small object. In the EOB theory, this spin of the effective test particle is $\sim \mu a /M$ even if the small object does not rotate. The omission of this term will only induce a relative error of the Hamiltonian at least two orders lower than the mass ratio. Furthermore, for decoupling the equations of motion, we used the approximation $F(r,\theta) \approx F(r)$. This usually induces an error at $O(10^{-2})\nu$ order, even at the edge of the LSO, the error still $\lesssim 0.1\nu$. The analysis of these approximations performed here showed that the errors could be ignored for EMRIs due to the very small mass ratio  (see Table I). By encoding the mass-ratio correction in the Hamiltonian $H_{\rm NS}$, our expressions may be an improvement compared to the test-particle approximation. 

However, as stated in the previous sections, the EOB's description at the extreme-mass-ratio limit does not get guarantee. Considering the comparison of the ISCO shifts with the gravitational self-force has obvious deviation for the extreme spin cases, we can only state our detailed results qualitatively reveal the influences of mass-ratio on the conservative dynamics. Fortunately, our analytical technology presented in this paper can be used to any similar deformed Kerr metrics. Once there is an updated version of the EOB resummation approximation, then the EOB corrections at the extreme-mass-ratio limit are improved or guaranteed, our results can be easily updated too by replacing the EOB potentials, and then we can get an accurate analytical formalism with mass-ratio corrections for the EMRIs.    

One of the scientific targets of EMRIs is to detect the
spacetime geometry of a SMBH. For this target, an accurate and efficient waveform template is needed. However, this is still a challenge. The analytical orbital solution including the mass ratio, eccentricity, and orbital inclination given in this paper is more accurate than the test-particle model and more convenient than the original EOB equations for
inclined-eccentric orbits over a long-term evolution.

Due to the analytical frequencies, a combination with the frequency-domain Teukolsky equation \cite{Teukolsky} will be more convenient and can generate numerical waveforms. In the future, we will use the formalisms in this work to generate the orbital evolution and waveforms for EMRIs. 

\acknowledgments
This work is supported by NSFC No. 11773059, and we also appreciate the anonymous Referee’s suggestions about our work.




\bibliographystyle{unsrt}

\end{document}